\documentclass[aps, prx, amsmath, amssymb, reprint,floats,floatfix,superscriptaddress]{revtex4-2}

\usepackage{xcolor}

\usepackage{graphicx}

\usepackage[export]{adjustbox}

\usepackage[colorlinks]{hyperref}

\usepackage{bm}

\usepackage{braket}

\usepackage{enumerate}
\usepackage[caption=false]{subfig}
\usepackage{physics}
\def\bra#1{\mathinner{\langle{#1}|}}
\def\ket#1{\mathinner{|{#1}\rangle}}

\newcommand{\be}{\begin{eqnarray}}
\newcommand{\ee}{\end{eqnarray}}

\def\doubleover#1{\overline{\overline{#1}}}
\usepackage{accents}
\newlength{\dhatheight}

\begin{document}

\author{Chun Y. Leung} \affiliation{Department of Physics, Lancaster University, Lancaster LA1 4YB, United Kingdom}
\author{Dganit Meidan} \affiliation{Department of Physics, Ben-Gurion University of the Negev, Beer-Sheva 84105, Israel}\affiliation{Université Paris-Saclay, CNRS, Laboratoire de Physique des Solides, 91405, Orsay, France.}
\author{Alessandro Romito} \affiliation{Department of Physics, Lancaster University, Lancaster LA1 4YB, United Kingdom}

\title{Theory of free fermions dynamics under partial post-selected monitoring}

\begin{abstract}
    Monitored quantum systems undergo Measurement-induced Phase Transitions (MiPTs) stemming from the interplay between measurements and unitary dynamics. When the detector readout is post-selected to match a given value, the dynamics is generated by a Non-Hermitian Hamiltonian with  MiPTs characterized by different universal features. Here, we derive a \emph{partial post-selected} stochastic Schr\"odinger equation based on a microscopic description of continuous weak measurement. This formalism connects the monitored and post-selected dynamics to a broader family of stochastic evolution. We apply the formalism to a chain of free fermions subject to partial post-selected monitoring of local fermion parities. Within a 2-replica approach, we obtained an effective bosonized Hamiltonian in the strong post-selected limit.
    Using a renormalization group analysis, we find that the universality of the non-Hermitian MiPT is stable against a finite (weak) amount of stochasticity. We further show that the passage to the monitored universality occurs abruptly at finite partial post-selection, which we confirm from the numerical finite size scaling of the MiPT.
    Our approach establishes a way to study MiPTs for arbitrary subsets of quantum trajectories and provides a 
    potential route to tackle the experimental post-selected problem.
\end{abstract}

\maketitle

The field of entanglement dynamics in monitored many-body systems has recently emerged as a promising arena to explore universal collective phenomena far from equilibrium.
The underpinning physics stems from generic unitary dynamics, which builds entanglement between different parts of the system, and measurements, which disentangle and localize information. The interplay between the two leads to Measurement-induced Phase Transitions (MiPTs) between phases with different entanglement scaling. 
MiPTs have been originally discovered in random quantum circuits~\cite{skinner2019measurement,chan2019unitary,li2018quantum,szyniszewski2019entanglement}.  
Fueled by the experimental progress in the realization of quantum simulators, the field has then  
established unexpected connections with condensed matter physics, statistical mechanics, and the field of quantum information science~\cite{fisher2023random}, with initial evidence of MiPTs reported in recent experiments~\cite{koh2023measurement,google2023measurement,noel2022measurement} 
%Unitary dynamics coupling different part of the system builds entanglement across the system while measurement disentangles and localises information. At the crossing point of these two opposing dynamics, much attention has been devoted due to the discovery of an intriguing phenomenon bringing condensed matter physics and quantum information together: measurement-induced phase transition (MiPT). MiPT arises when the entangling unitary dynamics is incompatible with the measurement operators, resulting in a competition between entanglement enhancement and information localisation, and is characterised by entanglement transition from phases with different entanglement scaling. 
Quite generally, the hybrid unitary-measurement dynamics underpinning MiPT  fall into 2 classes: quantum circuits with unitary gates punctuated with measurement~\cite{fisher2023random}, 
%with some consisting of measurement-only dynamics(area law to area law transition), 
and system evolving under continuous measurements and Hamiltonian dynamics~\cite{poboiko2023theory,kells2023topological,carisch2023quantifying,coppola2022growth,coppola2022growth,cao2019entanglement,alberton2021entanglement,kells2023topological,buchhold2021effective,yamamoto2023localization,szyniszewski2022disordered,fava2023nonlinear,turkeshi2022enhanced,chahine2023entanglement,doggen2022generalized,jian2023fermion,xing2023interactions,tang2020measurement,buchhold2022revealing,ladewig2022monitored,Minoguchi2022CFT,muller2022measurement,jin2023measurement,jian2022criticality,zhang2022universal,botzung2021engineered}. 
In both scenarios, MiPTs between phases with distinct topological quantum order from measurements-only dynamics have also been identified~\cite{lavasani2021topological,kells2023topological,roy2020measurement,rossini2021coherent,sang2021measurement,ippoliti2021entanglement,lavasani2021measurement}.

Entanglement phase transitions can originate from a different kind of non-unitary dynamics generated by non-Hermitian Hamiltonians. 
In the simplest terms, non-Hermitian Hamiltonians describe dissipation and/or gain in a system providing one of the simplest ways to model non-equilibrium. 
These models  too have shown novel entanglement transitions~\cite{li2020critical,kawabata2023entanglement,turkeshi2023entanglement,gopalakrishnan2021entanglement,hamazaki2019non,mu2020emergent},  and transition between states with different topological order~\cite{zirnstein2021bulk,zirnstein2021exponentially,malzard2015topologically,herviou2019entanglement,leykam2017edge,yao2018edge}.
Notably, some non-Hermitian dynamics can be established as a limit of monitored systems when 
retaining a pre-determined measurement readout (full post-selection).
The post-selection limit is most easily seen as the no-click limit in the quantum jump process where one only post-selects quantum trajectories with no-click events~\cite{jacobs2014quantum}, but it is a generic feature of monitored dynamics~\cite{turkeshi2021measurement,turkeshi2023entanglement,kells2023topological,le2023volume,zerba2023measurement,stefanini2023orthogonality,paviglianiti2023enhanced,turkeshi2023entanglement,zerba2023measurement,turkeshi2022entanglement,feng2023absence,su2023dynamics,chen2020emergent,tang2021quantum,jian2021yang,despres2023breakdown}. 

\begin{figure}
    \centering
    \includegraphics[width=0.45\textwidth]{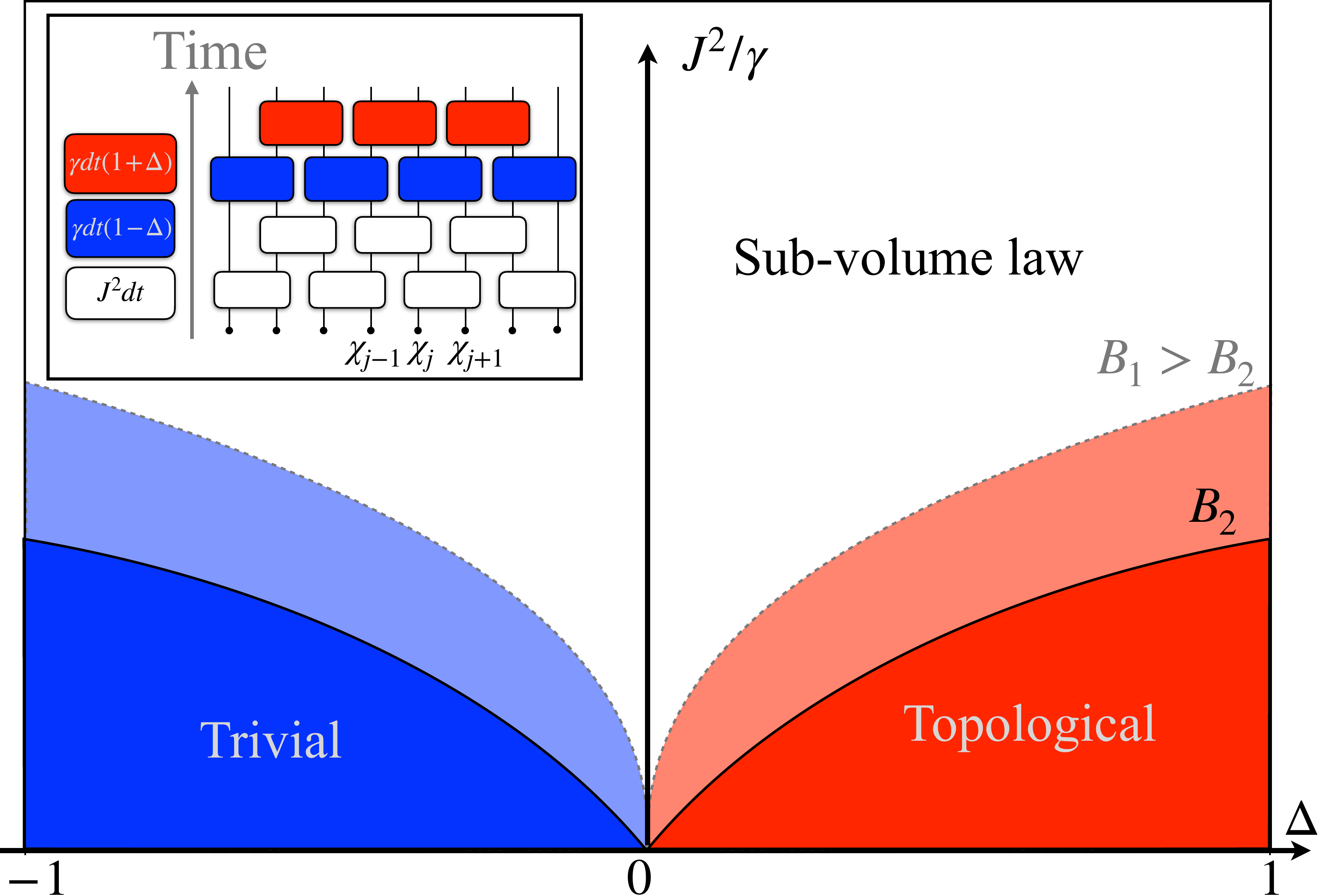}
    \caption{Schematic drawing of the phase diagram for the model in the inset under partial post-selection. Under strong partial post-selection, reducing the degree of post-selection $B1>B_2$, leads to an expansion of the sub-volume law phase.
    Inset: quantum circuit representation of the model consisting of random unitary evolution (white) and competing sets of parity measurements (blue and red) of Majorana fermions.}
    \label{fig:sketch of model and phase}
\end{figure}

MiPTs in the post-selected limit of monitored dynamics exhibit key differences compared to their monitored counterparts. These differences extend from features of the phase diagram to the universality class of the transition~\cite{kells2023topological,turkeshi2021measurement}. 
There have been some steps to incorporate sparse quantum jumps beyond the post-selected limit~\cite{zerba2023measurement,paviglianiti2023enhanced} or to map the full crossover explored numerically~\cite{kells2023topological},
%or in the averaged Lindblad dynamics, 
however, a theory that captures a systematic way to include a fraction of trajectories and explains the change in MiPTs properties is generally lacking. This is the question we address in this paper. 
We first summarize here our main findings.

We derive a \emph{partial-post-selected} (PPS) \emph{stochastic Schr\"{o}dinger equation} (SSE) --- cf. Eq.\eqref{eq:PPS SSE}, with a continuous parameter $B$ that controls the range of detector's outcomes that are retained. The PPS-SSE includes the fully monitored and fully post-selected dynamics as limiting cases and is valid for a generic quantum system with a continuously monitored Hermitian observable.

Next, we apply our analytic PPS approach to study the MiPT driven by non-commuting sets of local parity measurements in a real free fermionic chain.  In this model, the post-selected dynamics feature an area-to-area topological MiPT driven by the competing measurements, with a different critical exponent than its monitored analogue~\cite{kells2023topological}. 
Using the PPS-SSE approach and within a 2-replica approximation,  we obtain an effective description of the steady-state out-of-equilibrium phases in terms of an effective Hamiltonian---cf. Eqs.~(\ref{eq:2 replica effective majorana Hamiltonian},\ref{eq:bosonised Hamiltonian},\ref{eq:2 XXZ Ham}). 
% Within a 2-replica approximation and an RG analysis around the post-selected limit, we obtain an effective description of the steady-state out-of-equilibrium phases in terms of an effective Hamiltonian ---cf. Eqs.~(\ref{eq:2 replica effective majorana Hamiltonian},\ref{eq:bosonised Hamiltonian},\ref{eq:2 XXZ Ham}). 
From a Renormalization Group (RG) flow analysis, we find that the post-selected universal properties of the MiPT persist when one moves away from the post-selected limit by increasing the range of outcomes retained ---cf. Sec.~\ref{subsection: measurement only}, Fig.~\ref{fig:measurement only phases}. 

Our calculation further shows that the Luttinger parameter of the effective bosonized theory for strong post-selection diverges at a finite value of partial post-selection,$B$, which may indicate a phase transition driven by the stochasticity from quantum trajectories. This result is supported by numerical calculation which identifies non-monotonic behavior of the critical exponent at similar values of  $B$ ---cf. Fig.~\ref{fig:critical exponent vary B}. 

In the presence of unitary dynamics, the partial post-selected model features two distinct area law phases separated by a sub-volume law phase. We find that the sub-volume phase becomes increasingly stable upon moving away from the post-selected limit, as shown in Fig.~\ref{fig:sketch of model and phase} ---cf. Sec.~\ref{subsection:monitored-unitary}, Fig.~\ref{fig:phases from numeric and 2 replica cal for general}.
%We first derive a generic analytical approach to partial post-selection (PPS), and we than apply it to a specific model resulting in a stochastic Schr\"oding equation with the monitored and the post-selected as extreme ends of continuously connected dynamics. 
%We then use PPS to study the out-of-equilibrium order of monitored Gaussian fermions, for which the fully monitored dynamics is known analytically~\cite{fava2023nonlinear}. 
%We treat the problem analytically within a 2-replica approximation and find that the phase space and universality properties of the post-selected non-Hermitian dynamics survive a finite amount of stochastic components from measurement-induced dynamics before a change to the monitored universality over a relatively narrow region. The location of change is predicted analytically in the 2 replica calculations showing close agreement with numerical simulations. 
%Our findings open up a way to study the effect of stochastic quantum trajectories in general settings. Establishing a link between MiPT in monitored and non-Hermitian dynamics, our result suggests that the physics of the latter captures some features of the monitored systems that might survive in the average dynamics. 

The rest of the paper is structured as follows. %We first summarise our findings in Sec.~\ref{section:summary}. 
We develop the formalism of partial post-selection in Sec.~\ref{section:PPS} and extend it to the replica formalism in Sec.~\ref{section:MiPT and replica}. Sec.~\ref{section:Gaussian fermion} presents the model of interest, with the corresponding effective 2-replica description in Sec.~\ref{section: 2 replica} and the effective theory for the strong-post-selection regime in Sec.~\ref{section:bosonisation and RG}. The results are presented in Sec.~\ref{section:MiPT results} with a final discussion and conclusions in Sec.~\ref{section:conclusions}. Throughout this paper, we shall use the terms `post-selected' and `monitored' to indicate respectively the fully post-selected measurement dynamics and the fully stochastic continuous measurement where all readouts are retained respectively.

%\section{summary of results}
%\label{section:summary}

% \begin{enumerate}
% \item partial stochastic schrodinger.
%     \item measurement only: for a specific model post selected limit stable with random and agree with numerics.
%     \item indication of a phase transition with respect to randomness. 
%     luttinger parameter diverges, numerics change. 
%     \item with unitary dynamics - more stochastic increases the volume law. 
%     \item can be applied to any model
% \end{enumerate}

%In most literature, either one of the following limit is considered: the fully stochastic limit in which all measurement records are retained, or the fully post-selected non-Hermitian dynamics limit where only a predetermined set of results are kept. Here, we introduce \textit{partial-post selection}(PPS) which interpolate between the two limits.

\section{partial post-selection} \label{section:PPS}

We consider the dynamics of a continuously monitored quantum system whose evolution is described by the stochastic Schrödinger equation (SSE) 
\begin{align}
\label{eq:SSE1}
d\ket{\psi_{t} } =&  \left[ -idtH - dt\frac{\gamma}{2}\sum_j \left(\hat{O}_j-\langle \hat{O}_j\rangle\right)^2 \right. \nonumber \\ 
& \left. +\sum_j dW_j \left(\hat{O}_j-\langle \hat{O}_j\rangle\right)\right]\ket{\psi_t},
\end{align}
where $\ket{\psi_{t}}$ is the system's state at time $t$, $\hat{O}_j$ the set of observables being measured, and $H$ the system's Hamiltonian. To lighten the notation, we shall drop the hat above the measurement operator unless it is needed for clarity.
Eq. \eqref{eq:SSE1} is the Ito formulation of stochastic dynamics with $dW_j$ uncorrelated Gaussian-distributed stochastic increments with $\overline{dW_j dW_k}=\gamma dt \delta_{j,k}$, where $\gamma$ is the inverse measurement time at which typical stochastic realizations of the quantum trajectories are close to the observable's eigenvalue.

\begin{figure}
\includegraphics[width=0.45\textwidth]{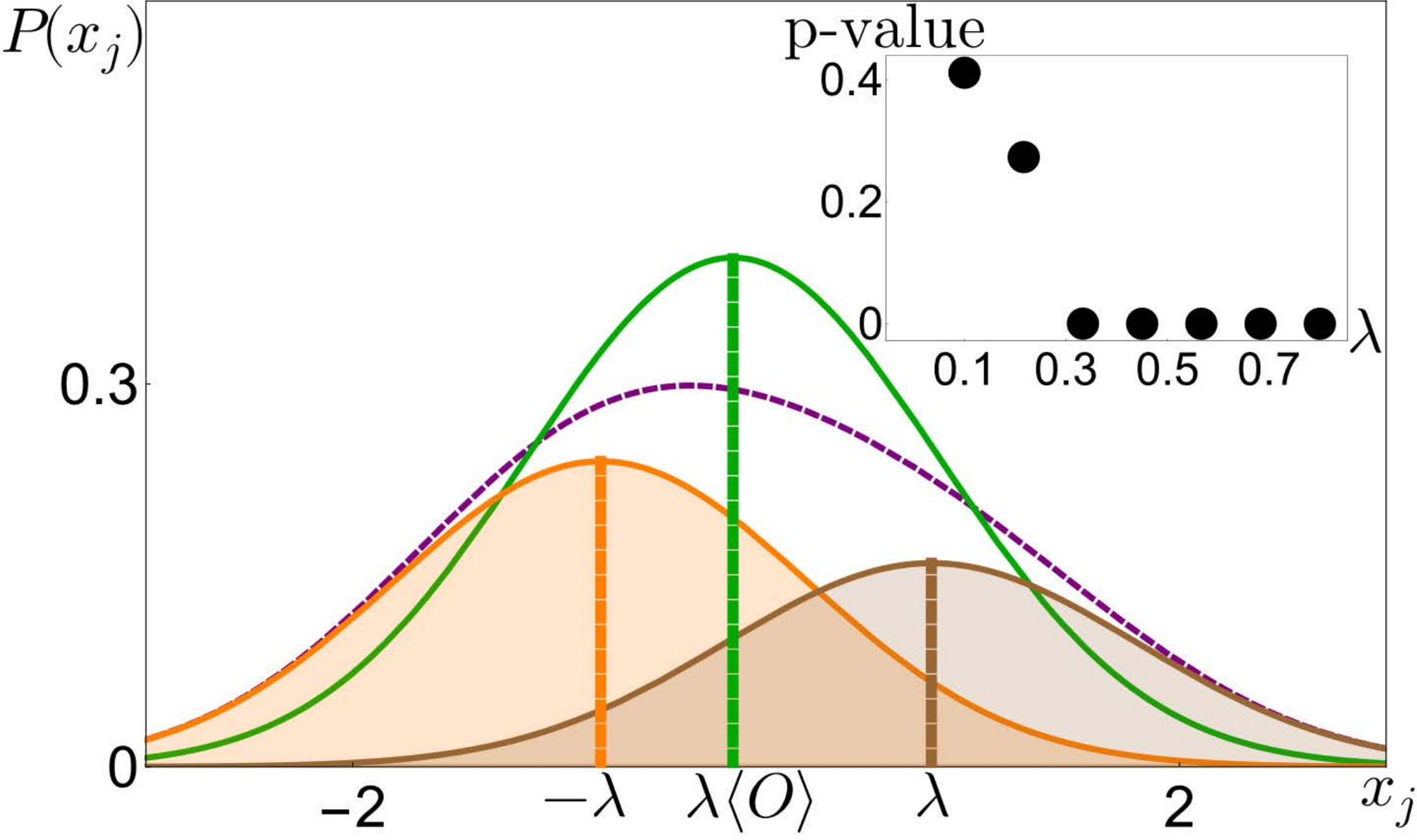}
\caption{Probability distribution of continuous Gaussian measurement readouts (\ref{eq:Prob dist continuous}). The readout distribution $P(x_j)$ (dashed purple) results from the sum of two overlapping Gaussians (brown and orange shaded),  centred at positions $\lambda$ and $-\lambda$ with different heights $\langle\Pi_{j,-}\rangle$ and $\langle\Pi_{j,+}\rangle$ respectively.
$P(x_j)$ is approximated by the Gaussian distribution  (green) in Eq.~(\ref{eq:Prob dist continuous}) which becomes exact in the limit of continuous measurements --- cf. inset. Inset: Accuracy of the approximation in Eq.(\ref{eq:Prob dist continuous}) quantified via a two-sample Kolmogorov-Smirnov test~\cite{sprent2007applied}. The accuracy (p-value) increases with decreasing $\lambda$, and becomes exact in the case of continuous measurement $\lambda\sim\sqrt{dt} \to 0$. 
The parameters are set as $\lambda=0.8$, $\langle\Pi_{j,+}\rangle=0.4$ and $\langle\Pi_{j,-}\rangle=0.6$. }
\label{fig:all prob dis}
\end{figure}

To develop the idea of partial post-selection, we start with a microscopic model of the measurement process leading to the SSE. 
We consider the measurement process described by a positively valued measurement~\cite{jacobs2014quantum}. In this case, after coupling the detector to the system in a state $\ket{\psi}_t$, the process returns a readout $x_j$, drawn from a probability distribution 
$
P(x_j)=\bra{\psi_t}K_j(x_j)^{\dagger}K_j(x_j)\ket{\psi_t}
$,
and a conditional state update $\ket{\psi_{t+dt}}=K_j(x_j)\ket{\psi_t}/\sqrt{P(x_j)}$. The process is entirely dictated by the Kraus operators $K_j(x_j)$.

We consider here specifically the case of continuously monitoring an observable $O_j$ as performed by a pointer with a continuous readout  $x_j$ with Gaussian a-priori distribution 
$G(x) = 1/\sqrt{2\pi\Delta^2}\exp(-x^2/2\Delta^2)$.
We further restrict to the simplest case of a measurement operator $\hat{O}_j=\hat{\Pi}_{j,+}-\hat{\Pi}_{j,-}$ which acts in the 2-dimensional space with projector $\hat{\Pi}_{+/-}$ and squares to identity $\hat{O}_j^2=\mathbb{I}$.
The Kraus operators are then
 given by~\cite{jacobs2014quantum}
\begin{equation}
\label{eq:Kraus for continuous}
    K_j(x_j,\lambda) = \sqrt{G(x_j-\lambda)} \Pi_{j,+} + \sqrt{ G(x_j+\lambda)} \Pi_{j,-},
\end{equation}
and 
\begin{equation}
\label{eq:Prob dist continuous}
P(x_j)= G(x-\lambda) \langle\Pi_{j,+}\rangle+ G (x+\lambda) \langle\Pi_{j,-}\rangle.
\end{equation}
The continuous SSE in Eq. \eqref{eq:SSE1} is recovered by setting $\lambda^2=\gamma dt$, where $dt \to 0$ with $\gamma$ finite guarantees $\lambda \ll 1$. In this limit, 
\begin{align}\label{eq:Kraus cont as single exp}
P(x_j)& \approx \frac{1}{\sqrt{2\pi\Delta^2}} \exp(-\frac{(x_j-\lambda \langle O_j\rangle)^2}{2\Delta^2}), \nonumber  \\
K_j(x_j,\lambda) & \approx \frac{1}{(2\pi\Delta^2)^{1/4}}\exp(-\frac{(x_j-\lambda O_j)^2}{4\Delta^2}).
\end{align}
The probability distribution is schematically shown in figure (\ref{fig:all prob dis}).
Notably, in eq. \ref{eq:Kraus cont as single exp}, we have used the fact that in the continuum limit $\lambda^2 =\gamma dt \to 0$, $\Delta\sim\mathcal{O}(dt^0)$.

The scenario of multiple measurement events can be written readily down: if there are $L$ lots of measurement operators $\hat{O}_j,j\in[1\dots L]$, the final state after measurements across all operators is 
\begin{align}\label{eq:many Kraus update}
    \ket{\psi_{t+dt}}=\frac{1}{\mathcal{N}}\prod_{j=1}^{L}K_j(x_j,\lambda)\ket{\psi_t},
\end{align}
where the results hold in the continuum limit $dt \to 0$ to order $\mathcal{O}(dt)$ also if some of the operators $\mathcal{O}_j$ do not commute. 
As a side note, (\ref{eq:Kraus for continuous}) can also be generalised to measurement operators with arbitrary spectrum with the same procedure illustrated above~\cite{jacobs2014quantum}.

The process of post-selection amounts to choosing and retaining the quantum trajectories that correspond to a  unique set of predetermined detector readouts $\{x_j\}$, while discarding the rest. 
We generalize this procedure to achieve Partial Post-Selection (PPS) by retaining all quantum trajectories that correspond to a finite range of detector outcomes.  
A natural means to achieve PPS is to force some degree of bias in the measurement outcome retaining the detector's outcome only if they are larger than a given, preset value, $r_c$. 
This amounts to truncating the readout probability distribution function $P(x_j)$ to a modified one,
\begin{align}\label{eq:modified Stat}
P_{r_c}(x_j)= P(x_j) \Theta(x_j-r_c) \approx e^{-\frac{(x_j-\lambda \langle O_j\rangle-\delta\lambda )^2}{2(\Delta+\delta)^2}} \equiv \underline{P} (x_j),
\end{align}
where $\Theta(x)$ is the Heaviside step function.

\begin{figure}
\includegraphics[width=0.45\textwidth]{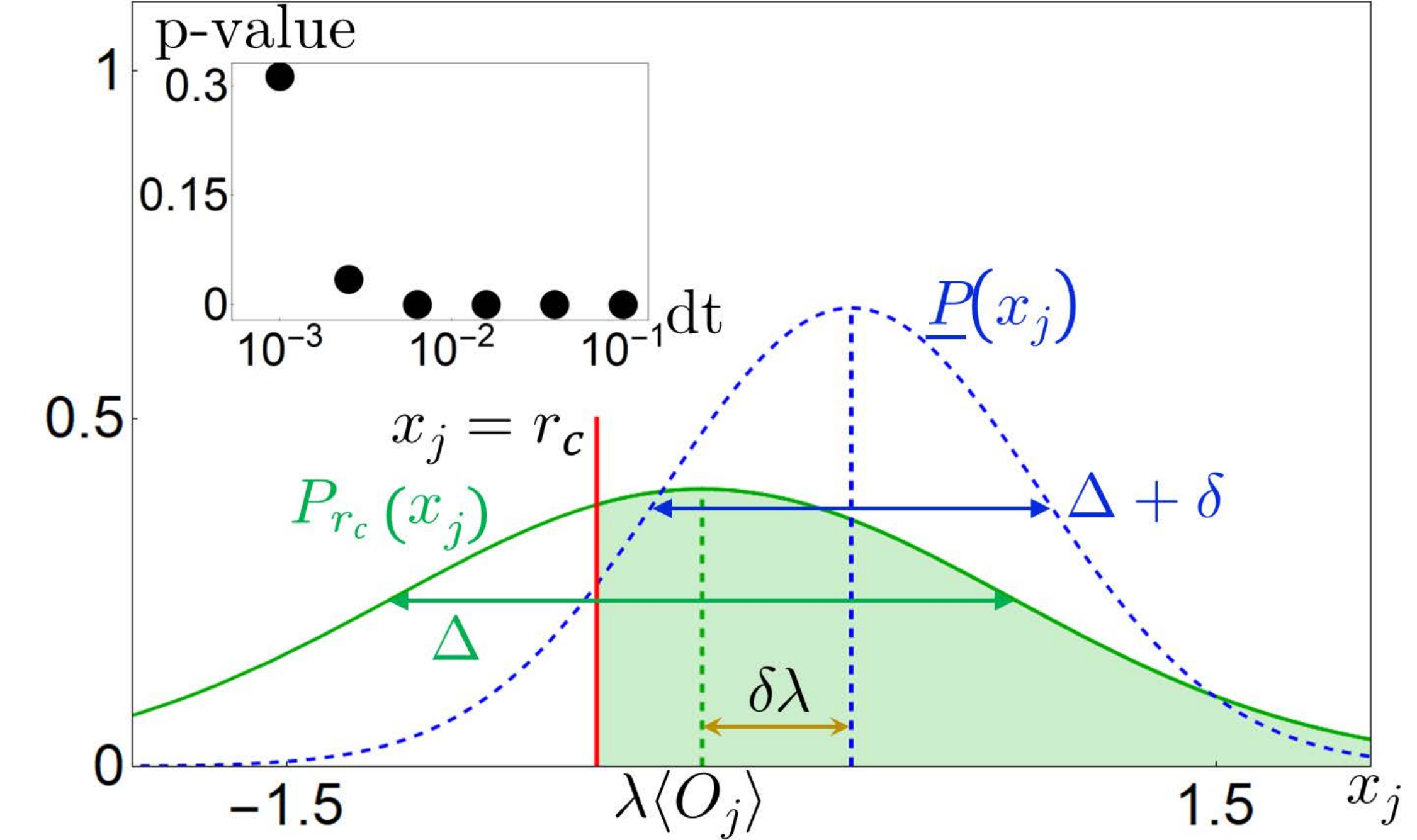}
\caption{Partial-post selection procedure in (\ref{eq:modified Stat}). The measurement outcome  Gaussian distribution (green) is truncated at $x_j=r_c$, resulting in a new distribution $P_{r_c}(x_j)$ (shaded) with shifted mean $\lambda\rightarrow\lambda+\delta\lambda$ and shifted variance $\Delta^2\rightarrow(\Delta+\delta)^2$. 
$P_{r_c}$  is approximated by a new Gaussian, $\underline{P}(x_j)$ (blue), with mean $\lambda+\delta\lambda$ and variance $(\Delta+\delta)^2$. The approximation is valid in the continuum limit as shown in the inset. Inset: p-value from a KS2 test for the two distributions $P_{r_c}$ and $\underline{P}$ with various $dt$. The approximation is exact in the continuum limit $dt \to 0$
The parameters are set as $\langle O_j\rangle=0.2,\ r_c=-0.5$ and $\lambda=0.3$.}
\label{fig:PPS drawing and 2 sam KS p value}
\end{figure}

In the last step in Eq. \eqref{eq:modified Stat}, we have approximated the truncated distribution by a Gaussian distribution whose mean and variance, parametrised by $\delta \lambda$ and $\delta$ respectively,  are determined by demanding that they coincide with those of $P_{r_c}(x_j)$, as illustrated in  Fig. (\ref{fig:PPS drawing and 2 sam KS p value}). 
While the distribution $P_{r_c}$ and $\underline{P}$ are generically different, we demand a proper scaling of $r_c$ with $dt \to 0$, so that the two distributions coincide in the continuum limit.  
This is achieved with the scaling 
\begin{align} \label{eq:b definition}
    \delta\lambda= b \lambda = b \sqrt{ \gamma dt},
\end{align}
where $b$ is kept constant in the limit $dt \to 0$ (see Appendix~\ref{sup:PPS and Gau}). 
The relation between $r_c$ and $b$ is derived and discussed in Appendix~\ref{sup:PPS and Gau}, and $b$ captures the discrete-time process $r_c$ in the time continuum limit, in analogy to $\gamma$ capturing the discrete process $\lambda$ in continuous measurement backaction. On the other hand, the correction in variance, $\delta$, can be safely ignored (Appendix~\ref{sup:PPS and Gau}). Importantly, at leading order in $dt$, the functional dependence of $r_c$ on $b$ is independent of the system's state, so that the continuum limit at constant $b$  corresponds to an operationally well-defined truncation of the probability $P(x_j)$.

We show explicitly via a two-sample Kolmogorov-Smirnov (KS2) test from a numerical sampling of $P_{r_c}$ and $\underline{P}$~\cite{sprent2007applied}, that the approximation by a Gaussian distribution in the time continuum analysis becomes exact in the continuum limit. The results are reported in figure \ref{fig:PPS drawing and 2 sam KS p value}, with the inset showing that the p-values (a statistical measure of overlap) of the two distributions are increasing with small time increments $dt$.

The continuum limit of $\underline{P}(x_j)$ in  Eq. \eqref{eq:modified Stat}, allows us to obtain a corresponding PPS SSE. Specifically, we introduce a new random variable $\xi_j=x_j/\Delta-\lambda\langle O_j\rangle-b\lambda \ , \ \textbf{mean}(\xi_j)=0 , \textbf{Var}(\xi_j)=1$. When expressed in terms of $\xi$, the update of the state by the Kraus operator in (\ref{eq:many Kraus update}) becomes
\begin{align}\label{eq:PPS Kraus update}
\ket{\psi_{t+dt}}&=\frac{1}{\mathcal{N}}\prod_{j}K_j(x_j,\lambda)\ket{\psi_t} \nonumber \\
    &=\frac{1}{\mathcal{N}_2}\prod_{j} e^{\left(\frac{\lambda^2(O_j-\langle O_j\rangle-b)^2}{4\Delta^2}+\xi_j\lambda\frac{O_j-\langle O_j\rangle-b}{2\Delta}\right)}\ket{\psi_t},
\end{align}
where overall factors have been reabsorbed in the state-normalization $\mathcal{N}_2$.
By setting $\Delta=1/2$, $\lambda^2=\gamma dt$ and noticing that the random variable $\xi_j\sqrt{\gamma dt}=dW_j$ fulfils $\overline{dW_j dW_k}=\gamma dt$, the state update in Eq. \eqref{eq:PPS Kraus update} defines a Wiener  process to order $dt$ and we arrive at the modified partial-post selected stochastic Schrödinger equation (PPS SSE)
\begin{align}\label{eq:PPS SSE}
 & d\ket{\psi_{t} } = -idtH \ket{\psi_t} - dt\frac{\gamma}{2}\sum_j \left(\hat{O}_j-\langle \hat{O}_j\rangle\right)^2 \ket{\psi_t} \nonumber \\
 & +Bdt\sum_j \left(\hat{O}_j-\langle \hat{O}_j\rangle\right) \ket{\psi_t} 
+\sum_j dW_j \left(\hat{O}_j-\langle \hat{O}_j\rangle\right)\ket{\psi_t},
\end{align}
where we set $B=b\gamma$ and $dW_jdW_k=\gamma dt\delta_{j,k}$.
Eq. \eqref{eq:PPS SSE}
is the first main result of our work. 
It generalises the SSE for observables of a 2-dimensional Hilbert space to account for a partial selection of trajectories defined in an operationally meaningful procedure. 
From, Eq. \eqref{eq:PPS SSE}, we can clearly identify two limits: for $\gamma\neq0$, $B=0$ we recover the standards SSE for monitored dynamics, while for $\gamma=0$, $B\neq0$ we are in the post-selected limit governed by a non-Hermitian Hamiltonian $H_{\text{eff}}=H+iB\sum_j\hat{O}_j$. 
We also note that the sign of the non-Hermitian term can be flipped by simply defining $r_c$ from the positive side of $x_j$. 
Note that the post-selected limit is operationally obtained by $\gamma \to 0$ but $b \propto 1/\gamma$ with $\gamma b = \textrm{const}$.

\section{Measurement induced transition and replicated dynamics}\label{section:MiPT and replica}

The effect of partial post-selection can equivalently 
be captured in the evolution of the system density matrix.
If we consider a density matrix evolved along a quantum trajectory with a certain set of measurement outcomes $\{x_t\},t\in[1...M]$ is a discrete-time index. The resultant density matrix is given by
\begin{align}\label{eq:traj un-norm density matrix}
    \rho_{\{x_t\}}=\frac{\check{\rho}_{\{x_t\}}}{\Tr[\check{\rho}_{\{x_t\}}]},
\end{align}
where $\check{\rho}_{\{x_t\}}=K_{x_M}\dots K_{x_2}K_{x_1}\rho_0K^{\dagger}_{x_1}K^{\dagger}_{x_2}\dots K^{\dagger}_{x_M}$ is the the \textit{un-normalised} density matrix along the trajectory,  $K_{x_l}$ the Kraus operator in Eq. \eqref{eq:Kraus for continuous} is associated with measurement outcome $x_l$ and $\rho_0$ is the initial \textit{normalised} density matrix. From here onward, we will specify an un-normalised density matrix by a caron above: $\rho=\check{\rho}/\Tr[\check{\rho}]$.

Since the probability of this trajectory, labelled by $\{x_t\}$, is $P(\{x_t\})=\text{Tr}[\check{\rho}_{\{x_t\}}]$, the average density matrix over all trajectories is
\begin{align}\label{eq:ave rho from unnorm rho}
    \overline{\rho}=\sum_{\{x_t\}} \frac{\check{\rho}_{\{x_t\}}}{\text{Tr}[\check{\rho}_{\{x_t\}}]}P(\{x_t\})=\sum_{\{x_t\}} \check{\rho}_{\{x_t\}},
\end{align}
where we use $\overline{\cdot}$ to indicate the average over quantum trajectories and the sum runs over all possible set of measurement outcome.
In the continuum limit, the evolution of $\overline{\rho}$ is governed by a Lindblad equation, which, in the typical cases of incompatible measurements and/or unitary dynamics considered here admits as a long-term fixed point the maximally mixed state $\lim_{t\to\infty}\overline{\rho}_t \sim\mathbb{I}$.

To capture non-trivial effects from measurements in the steady-state ensemble of quantum trajectories, one, therefore, needs to consider the evolution along individual trajectories, i.e. via post-selection, or resort to averages of quantities which are non-linear in density matrix e.g.$\overline{\langle\hat{O}\rangle^k}=\overline{\Tr[\hat{O}\rho]^k}$, or the $k$-th Reny\`i entropy $\overline{S_{k}}=1/(1-k)\log\left(\Tr[\rho^k]\right)$, $k>1$. 

To treat analytically these non-linear averages one can resort to a powerful mathematical construction, the replica trick~\cite{bao2021symmetry,fava2023nonlinear,poboiko2023theory,bao2020theory}, where one considers identical replicas of the system's density matrix. The average of the replicated density matrix can then be related to the average of the non-linear quantities we are interested in. 
In fact,
\begin{align}\label{eq:replica trick demo}
    \overline{\langle O\rangle^k}&=\sum_{\{x_t\}}\left(\text{Tr}[O\rho_{\{x_t\}}]\right)^k P(\{x_t\})\nonumber \\&=\sum_{\{x_t\}}\text{Tr}[O^{\otimes k}\check{\rho}^{\otimes k}_{\{x_t\}}](\text{Tr}[\check{\rho}_{\{x_t\}}])^{1-k} \nonumber \\&=\lim\limits_{n\to1} \sum_{\{x_t\}}\Tr\left[\left(O^{\otimes k }\otimes\mathbb{I}^{\otimes n-k}\right)\check{\rho}^{\otimes n}_{\{x_t\}}\right],
\end{align}
and the non-linear quantity encoding the non-trivial effects of measurement-induced dynamics is the trajectories averaged $n$-replicated un-normalised density matrix $\sum_{\{x_t\}}\check{\rho}^{\otimes n}_{\{x_t\}}=\overline{\check{\rho}^{\otimes n}},n\geq k$. Eq. \eqref{eq:replica trick demo} shows that the fundamental object of interest is $\overline{\check{\rho}^{\otimes n}}$, and the limit $n\rightarrow1$ is an analytical continuation for $k>1$. Importantly, the replica limit here is taken as $n\rightarrow1$, which poses a different case from the standard limit  $n\rightarrow0$ resulting from the replica trick for disordered systems~\cite{evers2008anderson,giamarchi2003quantum}.

Since the entanglement between two subsystems is determined by the purity of one of them, the simplest MiPT in the entanglement properties of the system can be identified by the $k$-th purity of a subsystem. The $k$-th purity $\mu_{k,\mathbf{A}}= \Tr[\rho_{\mathbf{A}}^k]$ of a subsystem $\mathbf{A}$ involves the $k$-th power of the reduced density matrix and can be calculated using the replica trick by introducing a suitable operator in the replica space, $\mathcal{C}_{k,\mathbf{A}}$, as~\cite{bao2021symmetry,fava2023nonlinear} 
\begin{align}\label{eq:subsystem purity}
    \mu_{k,\mathbf{A}}=\Tr[\rho_{\mathbf{A}}^k]=\Tr[\mathcal{C}_{k,\mathbf{A}}\rho^{\otimes k}],
\end{align}
where $\rho$ is the \emph{normalised} density matrix and
\begin{equation}\label{eq:C for matrix multi}
    \mathcal{C}_{k,\mathbf{A}} =\sum_{\mathbf{A}_j}\bigotimes^{j=k}_{j=1}\ket{\mathbf{A}_j}\bra{\mathbf{A}_{j+1}},
\end{equation}
and the sum indicates that $\ket{\mathbf{A}_j}, \ (\text{j mod k})$ runs over all the basis in the subsystem $\mathbf{A}$. Note that
$\mathcal{C}_{k,\mathbf{A}}$ acts as an identity outside of $\mathbf{A}$ while cyclically permutes kets from $k$ replicated density matrix in $\mathbf{A}$. The associated $k$-th Rényi entropy $S_{k,\mathbf{A}}$, which is an entanglement measure, is related by $S_{k,\mathbf{A}}=1/(1-k)\log\mu_{k,\mathbf{A}}$.

Throughout most of this paper, we focus on the 2-replica analysis
and replace the true replica limit $n\rightarrow 1$ by $n=2$. This 2-replica approximation captures non-trivial measurement-induced effect, includes information about the true replica limit, and hence provides a valid figure of merit to identify the non-trivial effects of PPS on measurement-induced dynamics~\cite{bao2020theory,bao2021symmetry}. The $2$-replica approximation allows one to access \textit{distorted} average of quantities with a quadratic dependence on the density matrix~\cite{bao2020theory,bao2021symmetry}, i.e. $k,n=2$ in Eq. \eqref{eq:replica trick demo} as
\begin{align}\label{eq:2nd distorted}
    \frac{ \sum_{\{x_t\}}\text{Tr}[O^{\otimes 2}\check{\rho}^{\otimes 2}_{\{x_t\}}]}{\sum_{\{x_t\}}\Tr[\check{\rho}^{\otimes 2}_{\{x_t\}}]}=\frac{\sum_{\{x_t\}}P(\{x_t\})^2\text{Tr}[O\rho_{x_t}]^2}{\sum_{\{x_t\}}P(\{x_t\})^2}=\overline{\overline{\langle O\rangle^2}}
\end{align}
% where the notation $\overline{\overline{\cdot}}$ has been introduced to indicate the average with the distorted probability and 
where the denominator is included for normalization. 
Eq. \eqref{eq:2nd distorted} shows that the approximation corresponds indeed to averaging with a \emph{distorted} probability distribution, now given by $P(\{x_t\})^2$, and we denote this by a double over line above, $\overline{\overline{\cdot}}$.

Within this approximation, the \emph{distorted} $2$-nd purity $\doubleover{\mu}_{2,\mathbf{A}}$ can be calculated using the operator in (\ref{eq:C for matrix multi}) as:
\begin{align}\label{eq:distorted purity}
    \doubleover{\mu}_{2,\mathbf{A}}=\frac{\Tr[\mathcal{C}_{k,\mathbf{A}}\sum_{\{x_t\}}\check{\rho}_{\{x_t\}}^{\otimes2}]}{\Tr[\sum_{\{x_t\}}\check{\rho}_{\{x_t\}}^{\otimes2}]},
\end{align}
and the associated conditional $2$-nd Rényi entropy $S_{k,\mathbf{A}}^{(\text{cond})}=-\log\doubleover{\mu}_{k,\mathbf{A}}$ is an entanglement measure of an extended system including both the system and the detector~\cite{bao2021symmetry,bao2020theory}. Thus, the trajectories averaged 2-replica $\sum_{\{x_t\}}\check{\rho}^{\otimes 2}_{\{x_t\}}$ shall capture MiPT in the system.

We also note that the calculation of entanglement entropy $S_{0,\mathbf{A}} \equiv \Tr[\rho_{\mathbf{A}}\log\rho_{\mathbf{A}}]$ for Gaussian fermions, at leading order, amounts to taking $k=2$ in \eqref{eq:replica trick demo}~\cite{klich2009quantum,chahine2023entanglement,poboiko2023theory}. Therefore in leading order, the 2-replica approximation is expected to retain some information about the true replica limit entanglement entropy.

In Sec. \ref{section: 2 replica}, we will show that the characterisation of the fundamental object in this 2 replica approximation $\lim\limits_{t\to\infty}\sum_{\{x_t\}}\check{\rho}^{\otimes 2}_{\{x_t\}}$, boils down to the study of a Hermitian Hamiltonian.

\subsection {Replica dynamics in PPS}\label{subsection:SSE to Gaussian}

In the case of continuous measurements, we are considering here, the equivalent of Eq. \eqref{eq:PPS SSE} for the density matrix along individual trajectory is given by the stochastic differential equation
\begin{align}\label{eq:PPS density matrix lindblad}
    \partial_t\rho&=-i\bigg[\left(H+iB\sum_j\hat{O}_j-\langle \hat{O}_j\rangle\right)\rho\nonumber \\
    &-\rho \left(H-iB\sum_j\hat{O}_j-\langle \hat{O}_j\rangle\right)\bigg]  \nonumber \\ &-\frac{\gamma}{2}\sum_j\comm{\hat{O}_j}{\comm{\hat{O}_j}{\rho}}+\sum_jdW_j\acomm{\hat{O}_j-\langle \hat{O}_j\rangle}{\rho}.
\end{align}
In this case, since the average over trajectories is, in fact, an average over the Gaussian-distributed measurement outcomes, it can be represented by an average over random non-Hermitian Gaussian noises, as shown in Ref.~\onlinecite{fava2023nonlinear}. 
Explicitly, we can rewrite the quantum trajectories average of an operator $\hat{O}$ in \eqref{eq:replica trick demo} as
\begin{align}\label{eq:Gaussian ave for monitoring}
\int_{\mathcal{A}_j(t_l)}\prod_{l=1}^{M}\mu\left(\mathcal{A}_j(t_l)\right)\text{Tr}[\hat{O}\check{\rho}^{\otimes n}_{\mathcal{A}_j(t_l)}]=\Tr[\hat{O}\mathbb{E}_G[ \check{\rho}^{\otimes n}_{\mathcal{A}_j(t_l)}]]
\end{align} 
and the notation $\mathbb{E}_G[\dots]$ indicates a Gaussian average over all random variables $\mathcal{A}_l$. In the monitored dynamics, the Gaussian measure $\mu(\mathcal{A}_j(t))$ has mean centred at $\mathbb{E}_G[\mathcal{A}_j(t)]=0$ and variance $\mathbb{E}_G[\mathcal{A}_j(t)\mathcal{A}_{j'}(t')]=\gamma\delta(t-t')\delta_{j,j'}$ in time continuum. 
The details of the derivation are summarised in Appendix \ref{sup:cont mea as nH and PPS}, where we follow the notation by Ref.~\onlinecite{fava2023nonlinear}. The result is a random non-Hermitian Hamiltonian acting on the \emph{un-normalised} density matirx, see Eq. \eqref{eq:supp con mea, rand nH Ham demo}.
The generalization to more than one set of measurements is straightforward, and here we abuse the notation $\mathbb{E}_G[\dots]$ to denote the Gaussian average over all random variables from all measurement processes, each with its Gaussian measure. 

This `non-Hermitian noise' formalism can be applied to the post-selection procedure in Sec. \ref{section:PPS}, which is formulated in terms of Gaussian distributed measurement readouts.
As shown in Eq. \eqref{eq:modified Stat}, the overall effect of PPS is shifting the centre of the Gaussian distribution of the measurement readouts by an amount $\delta \lambda=b\lambda$. 
When taking the continuum limit $dt \to 0$, the averages of stochastic process in the PPS Schr\"odinger equation \eqref{eq:PPS SSE} is equivalently described in the `non-Hermitian noise' formalism by a Gaussian distribution 
with a shifted mean of the measure $\mu\left(\mathcal{A}_j(t)\right)$ (see Appendix~\ref{sup:cont mea as nH and PPS} for the detailed derivation)
\begin{align}\label{eq:PPS on Gaussian ave}
    & \mathbb{E}^{(PPS)}_G[\mathcal{A}_j]=b\gamma=B, \nonumber \\ &\mathbb{E}^{(PPS)}_G[\mathcal{A}_j\mathcal{A}_k]=\gamma \delta(t-t')\delta_{j,k} + B^2. 
\end{align}
The procedure can be further extended to deal with averages in the replica formalism. The fundamental object of interest in the replica dynamics (cf. Eq. \eqref{eq:replica trick demo}) is then $\mathbb{E}_G[\check{\rho}^{\otimes n}_{\mathcal{A}}]$. We will show in Sec.\ref{section: 2 replica} that this will lead to an extra deterministic non-Hermitian term in the PPS dynamics.

\section{Gaussian fermion model}\label{section:Gaussian fermion}

We apply now the formalism of partial postselection to a specific model where the MiPT has been predicted~\cite{fava2023nonlinear}. The model, sketched in Fig.~\ref{fig:sketch of model and phase}, consists of a chain of real Majorana fermions with unitary dynamics governed by nearest-neighbour hopping random Gaussian white noise, and continuous weak monitoring of parity of nearest neighbours pairs of fermions. Within the Gaussian averaging formalism employed in Sec. \ref{subsection:SSE to Gaussian} (cf. Appendix \ref{sup:cont mea as nH and PPS}), the dynamics of the model are governed by a non-Hermitian random Hamiltonian given by (see also Eq. \eqref{eq:supp con mea, rand nH Ham demo})
\begin{align}\label{eq:OG majorana model}
    H(t)=\sum_j^{L}\left[J_j(t)+iM_{j}(t)\right]i\chi_j\chi_{j+1}
\end{align}
and $L$ (even) is the length of the chain which is always even. $J_j(t)$ and $M_{j}(t)$ are Gaussian random variables in space and time with 
\begin{align}\label{eq:J properties}
    \mathbb{E}_G[J_j(t)]=0, \, \mathbb{E}_G[J_j(t)J_{j'}(t')]=J^2\delta(t-t')\delta_{j,j'},
\end{align}
and the properties of  the non-Hermitian Gaussian noise $M_j(t)$ follow from the  Eq. \eqref{eq:PPS on Gaussian ave} to give
\begin{align}\label{eq:M properties}
    &\mathbb{E}_G[M_j(t)]=B_j, \, \nonumber \\ &\mathbb{E}_G[M_j(t)M_{j'}(t')]=\gamma_j\delta(t-t')\delta_{j,j'}+B_jB_{j'},
\end{align}
where the partial post-selection is now entirely controlled by $B$.
We further specify the measurement/PPS strength dependence on the individual sites to be
\begin{align}\label{eq:dimerise}
    \gamma_j&=\gamma(1+\Delta(-1)^j) \nonumber \\
B_j&=B(1+\Delta(-1)^j)
\end{align}
so that $-1\leq \Delta\leq1$ describes dimerisation in measurement/PPS strengths. 
This groups the measurement operators into two non-commuting (and competing) sets: the odd and even bonds parity measurement, each with measurement strength $\gamma(1-\Delta)$ and $\gamma(1+\Delta)$ respectively.

This model has been investigated in the monitored limit $B=0$ in Ref.~\onlinecite{fava2023nonlinear}. It was predicted to undergo  MiPTs between area and $log^2$-scaling entanglement entropy as a result of  the competition between unitary dynamics and measurement.  The measurement-only limit, $J=0$, of the model consisting of two sets of competing measurements coincides with the one investigated in Ref.~\onlinecite{kells2023topological}. The MiPT therein shows a peculiar dynamical critical exponent in the full monitored limit which differs from the projective counterpart (of a percolation universality class~\cite{lavasani2021topological,lavasani2021measurement}) and the fully-post-selected limit (of Ising universality class~\cite{kells2023topological}). 

We can now proceed with the analysis of the model according to the general formalism in Sec. \ref{sup:cont mea as nH and PPS}. 
The evolution of the unnormalised density matrix $\check{\rho}(t)$, is governed by Eq. \eqref{eq:traj un-norm density matrix}, which, in the time-continuous limit considered here reduces to
\begin{align}\label{eq:Kraus, dm evolve}
    K(t)&=\exp[-i\int_0^{t}dt'H(t')] \nonumber \\ \check{\rho}_{J,M}(t)&=K(t)\rho(0)K^{\dagger}(t), \ \rho_{J,M}(t)=\frac{\check{\rho}_{J,M}(t)}{\text{Tr}[\check{\rho}_{J,M}(t)]},
\end{align}
and we label the trajectories by $J$ and $M$, the set of random noise and measurement outcomes (see Appendix B for this time continuum process). 

To proceed further, it is advantageous to employ the Choi–Jamiołkowski isomorphism to map operators into states~\cite{jamiolkowski1972linear,choi1975completely}. In this way, we can first express the density matrix of $n$-replica Majorana chains of length $L$ as a state of $2n$ Majorana chains. The evolution operator then acts as a superoperator on the duplicated Hilbert space.
\begin{align}\label{eq:Choi Kraus}
    \check{\rho}^{\otimes n}(t)\xrightarrow[]{Choi}\ket{\check{\rho}^{\otimes n}(t)}\rangle=\left(K(t)\otimes K^*(t)\right)^{\otimes n}\ket{\rho^{\otimes n}(0)}\rangle
\end{align}
where the object $\ket{\dots}\rangle$  indicates that the states are in the duplicated Hilbert space. The details of the isomorphism and the derivation of Eq. \eqref{eq:Choi Kraus} are reported in appendix \ref{sup:Choi and replica trick}. In this state-formalism, the trajectory-averaged $n$-replicated un-normalised density matrix is given by 
\begin{align}\label{eq:Choi n replica}
    \mathbb{E}_G[\ket{\check{\rho}^{\otimes n}(t)}\rangle]&\equiv\ket{\check{\rho}^{(n)}(t)}\rangle \nonumber \\
    &=\mathbb{E}_G[\left(K(t)\otimes K^*(t)\right)^{\otimes n}]\ket{\rho^{(n)}(0)}\rangle,
\end{align}
and we shorthand $\ket{\check{\rho}^{(n)}(t)}\rangle$ for the average un-normalised $n$-replicated density matrix in the duplicated Hilbert space. In particular under Choi–Jamiołkowski isomorphism, the trace operation in Eq.  \eqref{eq:replica trick demo} becomes a transition amplitude 
\begin{align}\label{eq:Choi ave as trans}
    \lim\limits_{n\to1} \text{Tr}\left[O^{\otimes k }\otimes\mathbb{I}^{\otimes n-k}\mathbb{E}_G[\check{\rho}^{\otimes n}(t)]\right] =\lim\limits_{n\to1} \langle\langle\mathcal{O}_k|\check{\rho}^{(n)}(t)\rangle\rangle,
\end{align}
where the boundary bra in the duplicated Hilbert is
\begin{align}
    \ket{\mathcal{O}_k}\rangle=\left(O\otimes\mathbb{I}\right)^{\otimes k}\otimes\left(\mathbb{I}\otimes\mathbb{I}\right)^{\otimes n-k}\ket{\mathbb{I}}\rangle.
\end{align}
The associated objects relevant to 2-replica are the steady state distorted 2-nd subsystem purity, which in the duplicated Hilbert space appear as
\begin{align}\label{eq:distorted 2nd cumulant choi}
    \doubleover{\mu}_{2,\mathbf{A}}&=\langle\bra{\mathcal{C}_{2,\mathbf{A}}}\ket{\check{\rho}^{(2)}(t)}\rangle.
\end{align}

Eq. \eqref{eq:Choi ave as trans} and \eqref{eq:distorted 2nd cumulant choi} shows that a transition in the averaged replicated dynamics is directly reflected by a transition in the steady state of $\ket{\check{\rho}^{(n)}(t)}\rangle$. Thus, the identification and characterization of MiPT reduces to the study of $|\check{\rho}^{(n)}(t)\rangle\rangle$ in the steady state dynamics.

\section{Two-replica approach}\label{section: 2 replica}

Following Sec. \ref{section:MiPT and replica}, in the rest of the paper we will analyze the MiPT in the two-replica averaged dynamics. 
The quantity of interest is  now the 2-replica averaged state, governed by Eq. \eqref{eq:Choi n replica} with $n=2$, so that   
the evolution of $\ket{\check{\rho}^{(2)}(t)}\rangle$ becomes
\begin{align}\label{eq:2 replica dynamics}
    \ket{\check{\rho}^{(2)}(t)}\rangle=\mathbb{E}_G[\left(K(t)\otimes K^*(t)\right)^{\otimes 2}]\ket{\rho^{(2)}(0)}\rangle=e^{-\mathcal{H}t}\ket{\rho^{(2)}(0)}\rangle.
\end{align}
The effective Hamiltonian $\mathcal{H}$, obtained by Gaussian averaging (see Appendix \ref{sup:Choi and replica trick}) is given by
\begin{align}
    \mathcal{H}=&\sum_j \frac{J^2}{2}\left(\sum_{\substack{s=\uparrow, \, \downarrow\\a=1,2}} \mathcal{P}_{i,i+1}^{(sa)} \right)^2 - \sum_J \frac{\gamma_j}{2}\left(\sum_{\substack{s=\uparrow, \, \downarrow\\a=1,2}}s \mathcal{P}_{i,i+1}^{(sa)}\right)^2 \nonumber \\ 
    & -\sum_{\substack{s=\uparrow, \, \downarrow\\a=1,2}}\sum_js B_j \mathcal{P}_{i,i+1}^{(sa)},
\label{eq:2 replica effective majorana Hamiltonian}
\end{align}
where $\mathcal{P}_{i,i+1}^{(sa)}=i\chi_i^{(sa)}\chi_{i,i+1}^{(sa)}$ is the parity operator of the pair of Majorana fermions $\chi^{(s a)}_j$ and $\chi^{(s a)}_{j+1}$  in the replicated space, and $s=\uparrow, \, \downarrow$ labels the ket and bra space, $a=1,2$ labels the replica index. 
Note that these newly-introduced Majorana operators differ from the ones in (\ref{eq:OG majorana model}) by a Klein factor to ensure proper anti-commutation following the convention in Ref.~\onlinecite{bao2021symmetry,fava2023nonlinear}. The definition of $\chi_j^{(s a)}$ in terms of the original degrees of freedom is given in Appendix~\ref{sup:Choi and replica trick}.

From Eq. (\ref{eq:2 replica dynamics}), it can readily be seen that the replica dynamics is dictated by an imaginary time Hamiltonian, and thus $|\check{\rho}^{(2)}(t\rightarrow\infty)\rangle\rangle$ is determined by the low energy physics of $\mathcal{H}$. The problem then reduces to the study of phase transition in equilibrium low energy states of $\mathcal{H}$.

To this end, we note that $\mathcal{H}$ in Eq. \eqref{eq:2 replica effective majorana Hamiltonian} 
describes an interacting fermionic model with a global $O(2)\times O(2)$ symmetry. The two $O(2)$ symmetries are generated by the operators $\sum_j i\chi_j^{\uparrow1}\chi_j^{\uparrow2}$ and $\sum_j i\chi_j^{\downarrow1}\chi_j^{\downarrow2}$, and they correspond to rotation among $n$ Majorana operators within the ket ($s=\uparrow$) and bra ($s=\downarrow$) sector. In the absence of PPS, $B=0$, the global symmetry is larger with $O(2)\times O(2) \rtimes \mathbb{Z}_2$, and is further enlarged for measurement-only or unitary-only cases~\cite{bao2021symmetry,fava2023nonlinear}. 
The two $O(2)$ symmetries indicate two conserved $U(1)$ charges. These in turn, can be interpreted as  the conservation of fermion number of two distinct  fermions species given by
%allows to construct 2 species of fermions whose total number is conserved 
~\cite{bao2021symmetry}
\begin{align}\label{eq:complex fermion 2 replica}
    c^{\dagger}_{j,\uparrow}&=\frac{\chi_j^{(\uparrow1)}+i\chi_j^{(\uparrow2)}}{2}, \, \nonumber \\ c^{\dagger}_{j,\downarrow}&=\frac{\chi_j^{(\downarrow1)}-i\chi_j^{(\downarrow2)}}{2}
\end{align}
and the two conserved $U(1)$ charges appear explicitly as $\comm{\sum_jc^{\dagger}_{j,s}c_{j,s}}{\mathcal{H}}=0, \, \text{with} \,s=\uparrow \, \text{or} \,\downarrow$.

Expressing the Hamiltonian in eq.~\eqref{eq:2 replica effective majorana Hamiltonian} in terms of these two fermion species, we arrive, after some algebraic manipulation, at the following spinful fermion Hamiltonian (detailed in appendix~\ref{sup:Bosonisation}) 
\begin{align}
    \mathcal{H} =&H_{umk}+H_{m}+H_{0}\nonumber \\
    H_{umk}= &\sum_j -4(\gamma_j+J^2)\sum_{s=\uparrow,\downarrow}(c^{\dagger}_{j,s}c_{j,s}-\frac{1}{2})(c^{\dagger}_{j+1,s}c_{j+1,s}-\frac{1}{2}) \nonumber \\
    H_m=& \sum_j4(\gamma_j-J^2)(c^{\dagger}_{j,\uparrow}c_{j+1,\uparrow}+c^{\dagger}_{j+1,\uparrow}c_{j,\uparrow})\times \nonumber \\
     &  \ \ \ \ \ \ ( c^{\dagger}_{j,\downarrow} c_{j+1,\downarrow} + c^{\dagger}_{j+1,\downarrow} c_{j,\downarrow}) \nonumber \\ 
    H_0=&-\sum_j2B_j\sum_{s=\uparrow,\downarrow}(c^{\dagger}_{j,s}c_{j+1,s}+c^{\dagger}_{j+1,s}c_{j,s}).
\label{eq:effective spinful fermion Hamiltonian}
\end{align}

The steady-state measurement-induced phase and their MiPT are fully determined by the ground-state properties and phases of $\mathcal{H}$ in Eq.~\eqref{eq:effective spinful fermion Hamiltonian}.
Importantly, $\mathcal{H}$ is number conserving in both spin-up and spin-down fermion species, and the long wavelength (low energy) physics of \eqref{eq:effective spinful fermion Hamiltonian} depends on the particle number, or, more precisely, on the filling factor.
The relevant particle number sector is fixed by \eqref{eq:distorted 2nd cumulant choi} to be the same as $\ket{\mathbb{I}} \rangle$, which is in the half-filling sector, as shown in Appendix \ref{sup:Bosonisation}.
%Moreover, the computation of the distorted half-cut purity via Eq. \eqref{eq:subsystem purity} leads to the same charge sector constrain due to the state $\ket{\mathcal{C}_{2,\mathbf{A}}}\rangle$ being in the same half-filling charge sector (cf. Appendix \ref{sup:Bosonisation}).
We therefore analyse the half-filling ground state of $\mathcal{H}$.

In the monitored limit of $B=0$, one takes the freedom to choose $\ket{\rho(0)}\rangle$ within the same representation as $\ket{\mathbb{I}} \rangle$ and $\ket{\mathcal{C}_{2,\mathbf{A}}}\rangle$, so that 
$\rho(0)\propto\mathbb{I}$, which allows one to obtain an exact solution~\cite{giamarchi2003quantum}.
With the condition $B=0$, the Hamiltonian has an enlarged symmetry, since the local total parity across all replica, $\mathcal{R}_j=\prod_{a=1}^2 i\chi_{j}^{(\uparrow a)}\chi_{j}^{(\downarrow a)}$ is conserved and the Hamiltonian is invariant under an extra global $\mathbb{Z}_2$ symmetry in the Choi space: $\chi_{j}^{(\uparrow a)}\longleftrightarrow\chi_{j}^{(\downarrow a)}$ (this generalises to $n$ replica as well~\cite{fava2023nonlinear}).
In this case, the Hamiltonian can be expressed entirely as a function of local $SO(4)$ generators written in Majorana operators,
\begin{align}
    S^{\alpha, \beta}_j=\frac{i}{2}\left[\chi_j^{\alpha},\chi_j^{\beta}\right],
\end{align}
and the states $\ket{\mathbb{I}} \rangle$ and $\ket{\mathcal{C}_{2,\mathbf{A}}}\rangle$ isolate the spin representation among different irreducible representations~\cite{fava2023nonlinear}. In Appendix \ref{sup:2 replica monitor}, we demonstrate an alternative way to obtain the exact solution where a mapping to an integrable model can be constructed via 2 different spin-$1/2$ operators analogous to the $\eta,\Sigma$ spin from the Hubbard model~\cite{yang1990so}. In the monitored case, we show that (\ref{eq:2 replica effective majorana Hamiltonian}) is equivalent to
\begin{align}\label{eq:2 XXZ Ham}
    \mathcal{H}\propto\sum_{\substack{\Theta=\Sigma,\eta\\j=1}}^L &\frac{1}{2}(1+\delta(-1)^j)\left[\Theta^+_j\Theta^-_{j+1}+\Theta^-_j\Theta^+_{j+1}\right] \nonumber\\ &+ J_{z,j}\Theta^z_j\Theta^z_{j+1},
\end{align}
where $\delta=\frac{\Delta\gamma}{16(J^2+\gamma)} \ , J_{z,j}=\frac{J^2-\gamma_j}{J^2+\gamma_j}$.
Eq. \eqref{eq:2 XXZ Ham} corresponds to 2 decoupled $XXZ$ spin-$1/2$ chains.

\section{Strong PPS and bosonisation}\label{section:bosonisation and RG}

An analytical solution of the ground state of Eq.~\eqref{eq:effective spinful fermion Hamiltonian} is not available. 
However, in the strong partial-post-selected limit, $B_j \gg \gamma_j,J^2$, and half-filling condition of interest here, the spectrum of excitation is approximately linear, and the problem can be treated within the standard abelian bosonization  procedure~\cite{giamarchi2003quantum}. This amounts to linearising the fermion operator around the Femri surface
\begin{align}
    c_{j,s}\propto e^{-ik_Fx_j}\tilde{\psi}_{\mathbf{R},s}(x_j) + e^{ik_Fx_j}\tilde{\psi}_{\mathbf{L},s}(x_j),
\end{align}
and introducing the bosonic fields $\theta_s$ and $\phi_s$ via
\begin{eqnarray}
    \tilde{\psi}_{\mathbf{L},s}(x) &\propto e^{i(\phi_{s}(x)+\theta_{s}(x))},\\
    \tilde{\psi}_{\mathbf{R},s}(x) &\propto e^{-i(\phi_{s}(x)-\theta_{s}(x))},
\end{eqnarray}
where $\tilde{\psi}_{\mathbf{L/R},s}(x)$ is the slowly varying part of left/right movers of the fermion~\cite{giamarchi2003quantum}. The low energy properties of $\mathcal{H}$ will then be described by the linearized bosonic Hamiltonian.

The full bosonization procedure for $\mathcal{H}$ is reported in appendix~\ref{sup:Bosonisation} which leads to the low energy effective Hamiltonian  
\begin{align}\label{eq:bosonised Hamiltonian}
    \mathcal{H}_\text{bos} & \approx\sum_{\epsilon=\sigma,\rho}\Bigg[\frac{1}{2\pi}\int_x u_{\epsilon}K_{\epsilon}(\nabla\theta_{\epsilon})^2+\frac{u_{\epsilon}}{K_{\epsilon}}(\nabla\phi_{\epsilon})^2 \Bigg] \nonumber \\
     & + \sum_{\epsilon=\sigma,\rho} \int_x\frac{2g_{\epsilon}}{(2\pi\alpha)^2}\cos(\sqrt{8}\phi_{\epsilon}) \nonumber \\
    & + \frac{2g_2}{(2\pi \alpha)^2}\int_x \sin(\sqrt{2}\phi_{\rho})\cos(\sqrt{2}\phi_{\sigma}),
\end{align}
where 
$\phi_{\rho}=\frac{\phi_{\uparrow}+\phi_{\downarrow}}{\sqrt{2}} \ \text{and} \ \phi_{\sigma}=\frac{\phi_{\uparrow}-\phi_{\downarrow}}{\sqrt{2}}$ are the charge and spin sectors fields. The coupling constants and Luttinger parameters are given by 
\begin{align}\label{eq:Luttinger parameter}
    & u_{\rho}K_{\rho} = u_{\sigma}K_{\sigma} \equiv v_F= 4B, \nonumber \\ 
    & \frac{u_{\rho}}{K_{\rho}}=v_F-\frac{32aJ^2}{\pi}, \nonumber \\
     & \frac{u_{\sigma}}{K_{\sigma}}=v_F-\frac{32a\gamma}{\pi}, \nonumber \\
     & g_{\rho}=-g_{\sigma}=-16(\gamma-J^2), \nonumber \\
     & g_2=16a\Delta(B\pi-\gamma),
\end{align}
where $\Delta$ is the dimerization, $v_F$ is the effective Fermi velocity in the non-interacting case (e.g. from $H_0$ in Eq.~\eqref{eq:effective spinful fermion Hamiltonian}), and $a$ the lattice constant can be set to unity, $a=1$.
In deriving Eq.~\eqref{eq:bosonised Hamiltonian}, we keep only the most relevant operator. In particular, we discard highly irrelevant (in the RG sense) terms $\propto \cos(4\phi_{\uparrow, \downarrow})$ originating from the {\it umklapp} terms in the Hamiltonian $H_{umk}$. We also retain only slow oscillating term with $0$-$k_F$ and $4$-$k_F$ components around the filling factor $k_F=\pi/2a$.

The expected validity of the bosonization treatment in the strong post-selected limit $B\gg J^2, \gamma$ is confirmed by Eq.~\eqref{eq:Luttinger parameter}. 
Indeed the charge and spin Luttinger parameters $K_{\rho},K_{\sigma}$ diverges at $J^2/B\equiv \mathcal{J}^{*2}= \pi/8$ and $\gamma/B\equiv \ \gamma^*=\pi/8$ respectively. 
Firstly, this constrains also other Luttinger parameters in Eq.\eqref{eq:Luttinger parameter}, so that the sign of $g_2$ is the same as  $\Delta$. 
Furthermore, the divergence indicates that the failure of bosonisation at those points may signify the onset of phase transition~\cite{giamarchi2003quantum}, analogous to the phase transition between XY-phase and ferromagnetic phase transition in XXZ spin 1/2 chain.
As discussed later in Sec.~\ref{section:MiPT results}, this indication is further confirmed by the numerical finite-size scaling results for measurement-only MiPT.

Within the bosonized theory in~\eqref{eq:bosonised Hamiltonian}, the ground-state phases of $\mathcal{H}_\text{bos}$ are obtained by the RG flow of the parameters $J^2$, $\gamma$ and $B$, which can be computed within standard methods~\cite{amit1980renormalisation,jose1977renormalization,giamarchi2003quantum}, noting that Eq.~\eqref{eq:bosonised Hamiltonian} is the Hamiltonian of a Sine-Gordon model~\cite{giamarchi2003quantum}.
Here we follow the procedures in~\cite{giamarchi2003quantum,jose1977renormalization} performing real space coarse-graining of the correlator of a pair of vertex operators i.e.$\langle\exp[-ia\phi(r_1)]\exp[-ia\phi(r_2)]\rangle_{\mathcal{H}}$. 
The details of the calculation are reported in Appendix \ref{sup:RG}. 
In the analysis below, we will separate the no-dimerization case $\Delta=0$, from the general case. In the former, $g_2=0$ identically (cf. Eq. \eqref{eq:Luttinger parameter}), so that it cannot be simply obtained as a limit of the general case for $\Delta \to 0$. For no-dimerization, $\Delta=0$, $g_{\rho}$ and $g_{\sigma}$ flows separately and the perturbative RG flow up to second order in $g_{\epsilon}$ and $K_{\epsilon}$ gives 
\begin{align}\label{eq:SG RG flow root 8}
    \partial_lK_{\epsilon}&=-\frac{y_{\epsilon}^2K_{\epsilon}^2}{2} \nonumber \\ \partial_ly_{\epsilon}&=(2-2K_{\epsilon})y_{\epsilon}, \nonumber \\
    y_{\epsilon}&=\frac{g_{\epsilon}}{\pi u_{\epsilon}} \ , \ \epsilon=\sigma,\rho
\end{align}
where $l$ is the logarithm of the RG time.  In the most crude analysis in first order of $g_{\epsilon}$, the coupling for $\cos\sqrt{8}\phi_{\epsilon}$ is irrelevant for the physically relevant scenario $K_{\epsilon}>1$. However, accounting for the flow for $K_{\epsilon}$ can result in one of the mode being gapped but not both simultaneously, as we numerically evaluate the RG flows.

For $\Delta>0$, the $g_2$ term is more relevant than the $g_{\epsilon}$ since the cosine of the former is with higher frequency, so we can safely discard the $\cos(\sqrt{8}\phi)$ terms in $\mathcal{\text{H}}$. 
The RG flow equations in this case are derived in Appendix~\ref{sup:RG} following a standard procedure, which leads to 
\begin{align}\label{eq:SG RG flow 2}
    \partial_lK_{\rho}&=-\frac{g_2^2K_{\rho}^2}{16\pi^2u_{\rho}^2}\frac{I(\mu_{\rho},K_{\sigma},\sqrt{2})}{2\pi}, \nonumber \\
    \partial_lK_{\sigma}&=-\frac{g_2^2K_{\sigma}^2}{16\pi^2u_{\sigma}^2}\frac{I(\mu_{\sigma},K_{\rho},\sqrt{2})}{2\pi},  \nonumber \\
    \partial_lg_2&=\Bigg(2-\frac{1}{2}(K_{\rho}+K_{\sigma})\Bigg)g_2,
\end{align}
where
\begin{align}\label{eq:RG coupled SG subtle}
    \left(\frac{u_{\sigma}}{u_{\rho}}\right)^2&=1+\mu_{\rho},  \nonumber \\
    \left(\frac{u_{\rho}}{u_{\sigma}}\right)^2&=1+\mu_{\sigma}\nonumber \\
    I(\mu,K,\beta)&=\int_{-\pi}^{\pi}d\theta (\frac{1}{1+\mu\mathrm{cos}\theta})^{\frac{\beta^2K}{4}}.
\end{align}
The RG flow in Eqs. (\ref{eq:RG coupled SG subtle},\ref{eq:SG RG flow root 8}) dictate the low energy physics of the model and are used in the next section to characterize the properties of the MiPT in the partial-post-selected model.

\section{measurement-induced phases and their transitions}\label{section:MiPT results}

We are now in the position to use the Hamiltonians (\ref{eq:2 XXZ Ham},\ref{eq:bosonised Hamiltonian}), along with the RG-flow equations (\ref{eq:SG RG flow root 8}, \ref{eq:SG RG flow 2}) to characterize the steady-state phases of partially post-selected dynamics of the Gaussian model in Eq. \eqref{eq:OG majorana model}.
We  study both the measurement-only dynamics ($J=0$) and unitary-measurement-induced phases ($J^2>0$), and we discuss them separately hereafter. 

\subsection{Measurement-only dynamics}\label{subsection: measurement only}

\begin{figure}
    \centering
    \includegraphics[width=0.45\textwidth]{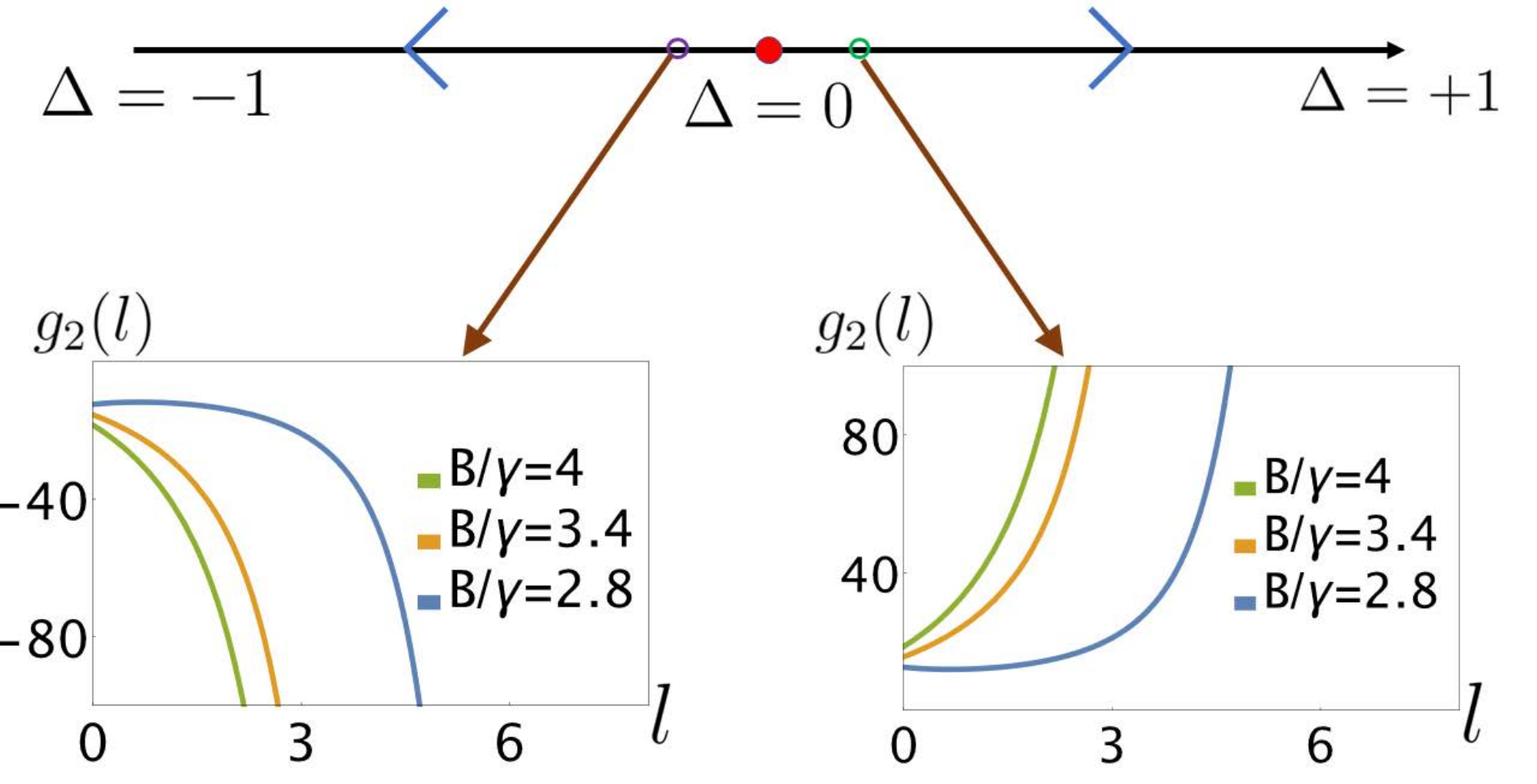}
    \caption{Schematic phase diagram for the measurement-only dynamics determined by the dimerization $\Delta$. The arrows indicate the RG flow to the two different fixed points $\Delta=\pm1$ with a critical point (red dot) at $\Delta=0$. 
    The plots show The RG flow of $g_2$ from (\ref{eq:SG RG flow 2}) evaluated  at different points (green and purple circles) indicating that the interaction is relevant in both cases, and it leads to area law of phases.}
    \label{fig:measurement only phases}
\end{figure}

In the absence of unitary dynamics, $J=0$, the system is evolving entirely according to two competing sets of measurements: the set of odd and the set of even bond measurements.
Notably, the $J=0$ limit of the model  \eqref{eq:OG majorana model} coincides with the measurement-only case studied in Refs.~\onlinecite{kells2023topological,fava2023nonlinear} where monitored and post-selected limits follow very different behaviours. In particular, finite-size scaling reveals that the monitored system belongs to a different universality class from the fully post-selected model~\cite{kells2023topological}. 

\paragraph{Post-selected limit---} The fully post-selected dynamics are obtained by setting $\gamma=0, \, B>0$ in Eq. \eqref{eq:2 replica effective majorana Hamiltonian}. In this case, $B$ is just an overall time scale and the physics is entirely dictated by $\Delta$, given by \eqref{eq:dimerise}. The effective Hamiltonian now reads:
\begin{align}\label{eq:post-selected Ham}
    \mathcal{H}=-\sum_{\substack{s=\uparrow, \, \downarrow\\a=1,2}}\sum_js B(1+\Delta(-1)^j) i\chi^{(s a)}_j\chi^{(s a)}_{j+1},
\end{align}
Using the usual Jordan-Wigner transformation, the imaginary time evolution is, therefore, equivalent to $2$ decoupled $1D$ traverse field Ising models  in either the bra ($s=\downarrow$) or ket ($s=\uparrow$) space.

The critical properties of the post-selected dynamics fall in the Ising universality class, and the critical exponent $\nu$ that determines the divergence of the correlation length $\xi\sim|\Delta|^{-\nu}$, is $\nu=1$.
Away from criticality for $\Delta>0$, the even parities are measured more strongly, and this phase is characterised by a pair of entangled Majorana fermions residing at the edges of a finite-length chain. 
This phase is associated with a $\log_2$ topological entanglement entropy per replicated chain in the area-law phase.

On the other side $\Delta<0$, the odd parities measurements are stronger and it features all Majorana being measured in pairs. This, therefore, corresponds to a topologically trivial phase with vanishing topological entanglement entropy.

\paragraph{Strong post-selection ---} 

When $\gamma \neq 0 $, the system does not follow any longer the deterministic dynamics from Eq. \eqref{eq:post-selected Ham}, but stochastic fluctuations inherent to the measurement process enter the system dynamics. With the partial post-selection introduced in Eq.~\eqref{eq:PPS SSE}, the parameter $\gamma/B$ controls the amount of fluctuations (i.e. the fraction of quantum trajectories) allowed in the system's dynamics. 
We can analyse the strong post-selected limit $\gamma/B \ll 1$ with the bosonized Hamiltonian \eqref{eq:bosonised Hamiltonian}. 
As argued in Sec. \ref{section:bosonisation and RG}, the steady state of the system is governed by different equations for $\Delta=0$ and $\Delta \neq 0$, so we address them separately.

For $|\Delta|>0$, using the flow in (\ref{eq:SG RG flow 2}), we observe that the $\cos$ operator corresponding to the $g_2$ coupling is in general relevant for $K_{\rho}+K_{\sigma}<4$, which we confirm by evaluating Eq. \eqref{eq:SG RG flow 2} numerically. 
The results are shown in Fig.~\ref{fig:measurement only phases} for different values of $\gamma/B$. 
The flow in the massive/gapped phase indicates an unbounded growth in the coupling $g_2$, which, moreover, does not change sign along the RG flow. 
When reinterpreting the RG flowing parameters in terms of the original parameters of the model $B$, $\gamma$ and $\Delta$ via~\eqref{eq:Luttinger parameter}, this limit  approaches the post-selected Hamiltonian \eqref{eq:post-selected Ham} ($\gamma \to 0$).  This indicates that the strong-PPS phase at finite $\gamma$ with
 $\Delta>0$  ($\Delta<0$) is continuously connected to the  gapped phase $\gamma =0,\Delta>0$ ($\gamma =0,\Delta<0$) of the post-selected model. The points $|\Delta|=1$ are the two only stable fixed points in the measurement-only dynamics, as reported in the phase diagram in Fig.~\ref{fig:measurement only phases}.
We therefore expect that the universal properties of the strong-partial post-selected regime are inherited from Eq.~\eqref{eq:post-selected Ham}, i.e. those of two uncorrelated copies of an Ising model.

This is the first main prediction of our theory: The MiPT remains in the same Ising-like university class for finite $\gamma/B$ as long as the bosonized approximation for the theory remains valid. 
Physically, this predicts the stability of the post-selected MiPT universal feature against (weak) fluctuations induced by the stochasticity of the measurement.

For $\Delta=0$, $J=0$, Eq. \eqref{eq:SG RG flow 2}, together with the definition of Luttinger parameters in Eq. \eqref{eq:Luttinger parameter}, implies that the $g_2=0$, and that the $\sigma$- and $\rho$-modes decouple. 
The RG-flow is then standard~\cite{giamarchi2003quantum}, with the $\rho$-mode flowing to a massive phase ($g_\rho \to \infty$), while the $\sigma$-mode, following an expansion around $K_\sigma \to 1^+$, flows to $g_\sigma \to 0$, $K_\sigma>1$.  
Correlations in the overall theory are thus dominated by the $\sigma$-mode, which is a Gaussian-free theory displaying free Luttinger liquid criticality, with a logarithmic scaling of the entanglement entropy.
Given that both  $\sigma$- and $\rho$-modes are massive for $\Delta \neq 0$,  the $\rho$-mode remains massive from $\Delta>0$ to $\Delta<0$ as long as $\gamma \neq 0$, while  the $\sigma$-mode undergoes a transition. This is compatible with the behaviour of an Ising-like transition as discussed above.

\paragraph{From strong post-selection to monitored dynamics---}
While our analytic theory predicts an Ising-like transition in the strong post-selected regime, numerical analysis indicates that the monitoring dynamics undergo a measurement-only transition of a different universality class with a critical exponent of $\nu=5/3$~\cite{kells2023topological}.
While our bosonized theory cannot access the full transition between post-selected ($\gamma/B=0$) and monitored ($\gamma/B \to \infty$) dynamics due to the breakdown of bosonization at $\gamma^*$(cf. Sec.~\ref{section:bosonisation and RG}), this divergence indicates a phase transition to a different phase~\cite{giamarchi2003quantum}.

\begin{figure}
    \centering
    \includegraphics[width=0.45\textwidth]{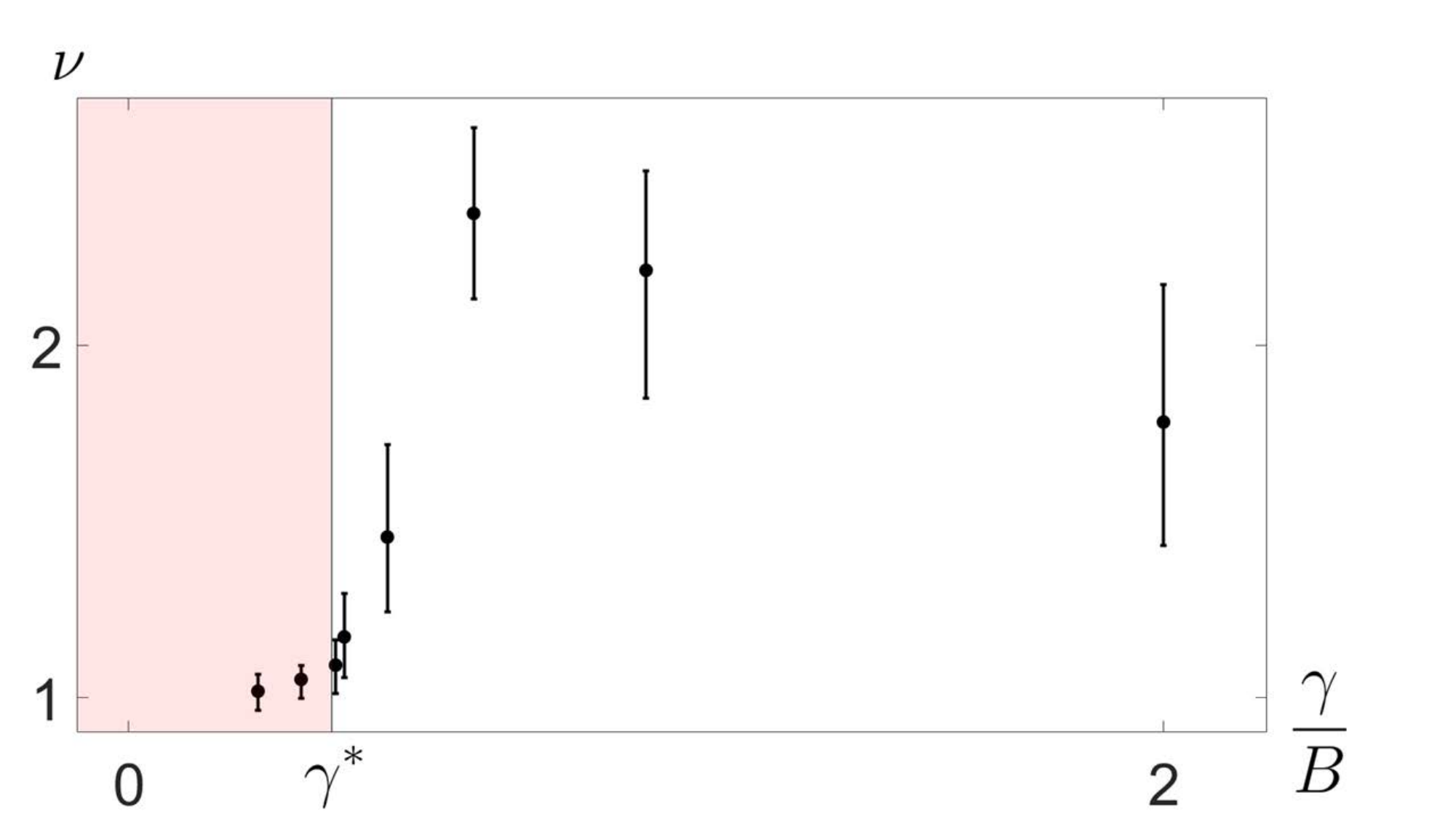}
    \caption{The critical exponent $\nu$ of the measurement only phase transition as a function of the degrees of PPS, $\gamma/B$. The pink region indicates where bosonization analysis is valid. The fully post-selected Ising critical exponent $\nu=1$ is unchanged for a finite range of $\gamma/B$. The abrupt deviation form $\nu=1$ occurs in the proximity of the breakdown of the bosonized theory at $\gamma/B \approx \gamma^* = 0.39$ (end of shaded region). The fully monitored critical exponent $\nu=5/3$ is recovered only for $\gamma/B \to \infty$.
    The large error bars for $\gamma/B \gg 1$ are due to the large fluctuation due to the increasing trajectory-to-trajectory fluctuations in this regime.}
\label{fig:critical exponent vary B}
\end{figure}

This observation is corroborated numerically. 
We analyze the critical exponent $\nu: \xi\sim |\Delta|^{-\nu}$ of this measurement-only MiPT via finite-size scaling for generic $\gamma/B$. 
The details of the numerical methods are reported in Appendix~\ref{app:numerics}. 
The results are presented in Fig. \ref{fig:critical exponent vary B} showing that $\nu \approx 1$ for strong PPS before deviating abruptly in a narrow range around $\gamma/B \approx \gamma^*$ and approaching  $\nu=5/3$ when  $\gamma/B \gg 1$.
Surprisingly, numerical data shows that close to the transition, $\nu \approx 2.3>5/3$, before dropping back to $\nu=5/3$ for larger $\gamma/B$.

\paragraph{Monitored limit, $B=0$ ---}
The monitored limit ($B=0$) is given by Eq.~\eqref{eq:2 XXZ Ham}, which, for the measurement only-case $J=0$, reduces to a $XXZ$-Hamiltonian with a dimerization in hopping term, also known as spin-Peierls model~\cite{giamarchi2003quantum}. 
This model predicts a BKT transition at $\Delta=0$~\cite{giamarchi2003quantum}. 
This differs from the Ising bosonized theory for the strong PPS. 
This  difference is also consistent with the change in the symmetry of the model in the two limits, as discussed in Sec~\ref{section: 2 replica}.

Note that the BKT transition (hence the scaling of the distorted partial purity or entanglement entropy) predicted by Eq.~\eqref{eq:2 XXZ Ham} does not capture  the correct universality class of the fully monitored dynamics. Indeed,  in the limit $B=0$, it has been shown that the $2$-replica model differs from the $n \to 1$ limit for which the phase transition in the measurement-only limit is not known~\cite{fava2023nonlinear}. 
However, since for strong PPS the replicas completely decouple, the post-selected limit is independent of the replica number, and we expect that the stability of the post-selected phase and its breakdown should be captured in the 2-replica case considered here.

\subsection{Partial post-selected monitoring with unitary dynamics}\label{subsection:monitored-unitary}

\paragraph{No-dimerization case, $\Delta=0$---} 
To analyze the effect of unitary dynamics on the system, we start by considering the case where dimerization is absent, $\Delta=0, J^2>0, \gamma>0$. 
In this case,  the RG flow in Eq.~\eqref{eq:SG RG flow 2} keeps $g_2=0$, the $\rho$-mode and $\sigma$-mode decouple as indicated by $H_{bos}$ and Eq.~\eqref{eq:SG RG flow root 8}, and the Luttinger parameters $K_{\rho}$ and $K_{\rho}$ in (\ref{eq:Luttinger parameter}) are  both initially larger than unity. 
At the leading order, the RG flow signals that $\mathcal{H}_\text{bos}$ is gapless for $K_{\epsilon}>1$. Evaluating (\ref{eq:SG RG flow root 8}) numerically reveals that one of the sectors is always massless. 
Given the decoupling between the two sectors, in general $\lim\limits_{t\to\infty}\ket{\check{\rho}^{(2)}(t)}\rangle$ will evolve to a tensor product of two ground states in the two sectors $\ket{\mathcal{H}_{\rho}}\otimes\ket{\mathcal{H}_{\sigma}}$. 
Thus, correlations w.r.t. $\mathcal{H}_\text{bos}$ are dominated by the gapless power-law decaying sector,
and the distorted partial purity $\doubleover{\mu}_{k,\mathbf{A}}=\langle\bra{\mathcal{C}_{2,\mathbf{A}}}e^{-\mathcal{H}t}\ket{\rho^{(2)}(0)}\rangle=\langle\bra{\mathbb{I}}\mathcal{C}_{2,\mathbf{A}}e^{-(\mathcal{H_{\rho}}+\mathcal{H_{\sigma}})t}\ket{\mathbb{I}}\rangle$ will be dominated by correlation in the gapless sector. 
Therefore, we expect that the entanglement entropy will be dominated by the critical sector and will show a critical entanglement scaling.

This result differs from the predicted $(\text{log}L)^2$ in Ref.~\onlinecite{fava2023nonlinear} for the fully monitored case.
% Note that for the fully monitored limit of our theory for the partial purity in the 2-replica approximation,  also predicts a $\log$-scaling critical phase associated with the inherent BKT transition. 
% However, as discussed in Sec.~\ref{subsection: measurement only}, the $2$-replica limit is not expected to capture the correct behaviour in this case~\cite{fava2023nonlinear}, while we expect the 2-replica results to be valid in the strong post-selected limit.
The absence of $(\text{log}L)^2$ scaling in strong PPS where bosonisation remains valid and could be traced back to the breaking of local parity $\mathcal{R}_j=\prod_{a}i\chi^{(+a)}_j\chi^{(-a)}_j,\comm{\mathcal{R}_j}{\mathcal{H}}\neq0$, which prohibits one to express $\mathcal{H}$ solely as local SO(2N) generators. Consequently, $\mathcal{H}$ is no longer described by the non-linear sigma model in~\cite{fava2023nonlinear} that gives the $(\text{log}L)^2$ scaling.

\begin{figure}
    \centering
    \includegraphics[width=0.45\textwidth]{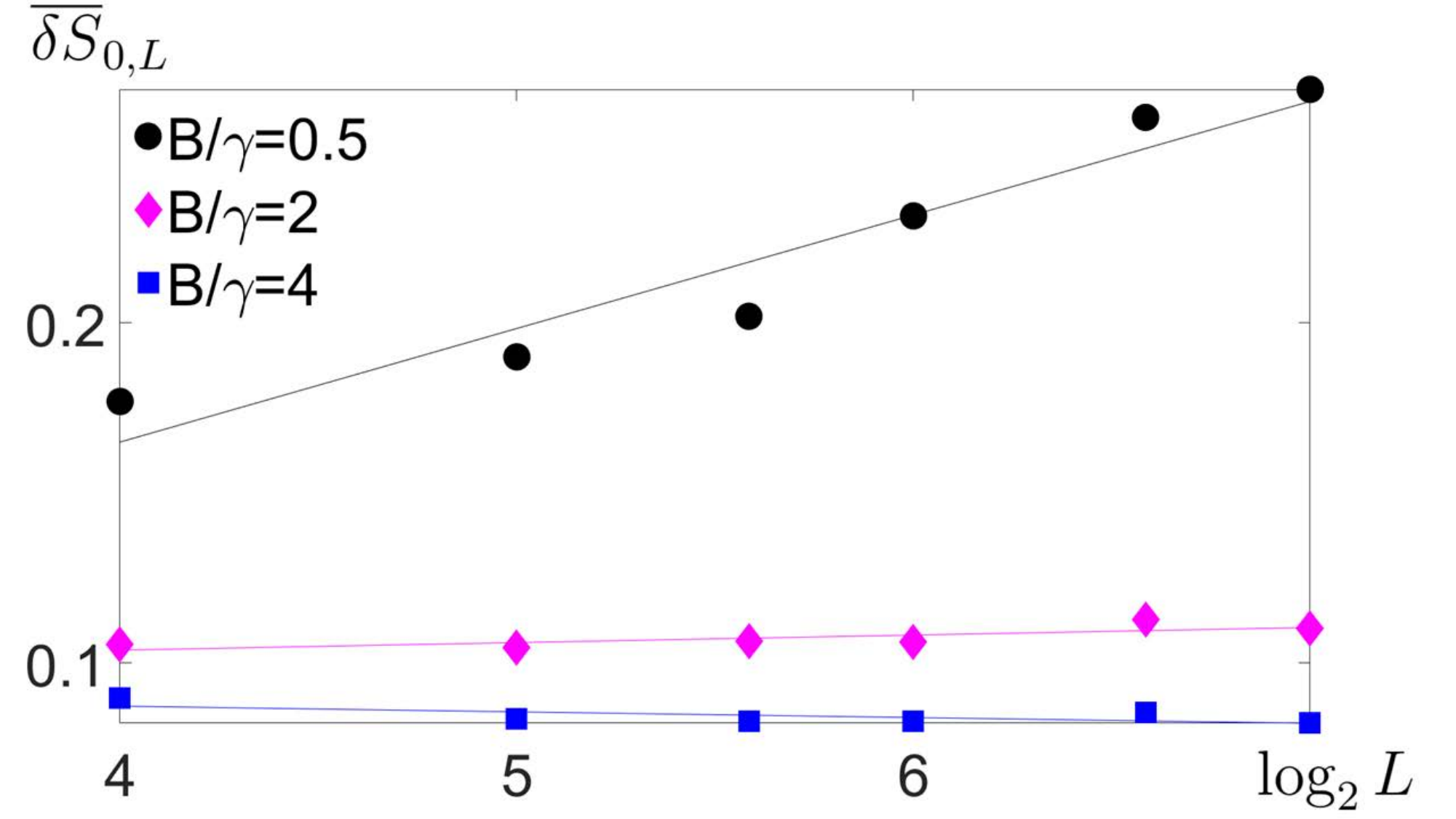}
    \caption{Scaling of the average entanglement entropy difference as a function of $\log L$ for  $\Delta=0$ and $J^2/\gamma=0.25$. 
    The dependence $\delta \overline{S}_{0,L}\sim\text{log}L$ or $\delta \overline{S}_{0,L}\sim\text{const.}$ distinguishes a $(\text{log}L)^2$ from log$L$ scaling respectively. The lines of best fit of the data indicate a change of $\delta \overline{S}_{0,L}$ from linear increase (pink and black) to constant (blue) with increasing $B/\gamma$.}
    \label{fig:0 dimerisation}
\end{figure}

To confirm a change from $ \log L$ to $ (\log L)^2$ with increasing $ J^2$, we numerically analyze the scaling of the entropy along the no-dimerization line. The results are shown in Fig.~\ref{fig:0 dimerisation}. 
To distinguish the log$L$-scaling from the $(\text{log}L)^2$ one, we use as the indicator the difference in half-cut entanglement entropy $\delta S_{0,L}\equiv S_{0,L_2/2}-S_{0,L/2}$ ($L_2=2L$)~\cite{fava2023nonlinear}, with $\delta \overline{S}_{0,L}\sim\text{log}L$ vs $\delta \overline{S}_{0,L}\sim\text{const.}$ dependence in the two cases respectively (see Appendix~\ref{app:numerics}). 
This analysis shows that, for fixed $\gamma$, increasing the degree of partial post-selection, $B$, leads to a change from $(\text{log})^2$-scaling to $\log$-scaling. 
The change in the scaling behaviour happens approximately at the point where bosonization is expected to break down $\mathcal{\gamma}^{*} \approx0.39$.
This is consistent with the picture in the previous section where the breakdown of bosonization at $\gamma*$  signals a transition away from the university of the post selected model towards the universality of the monitored model which is captured by the non-linear sigma in Ref.~\onlinecite{fava2023nonlinear}.

\paragraph{General monitored-unitary dynamics, $\Delta \neq 0$ ---}

\begin{figure}
\includegraphics[width=0.45\textwidth]{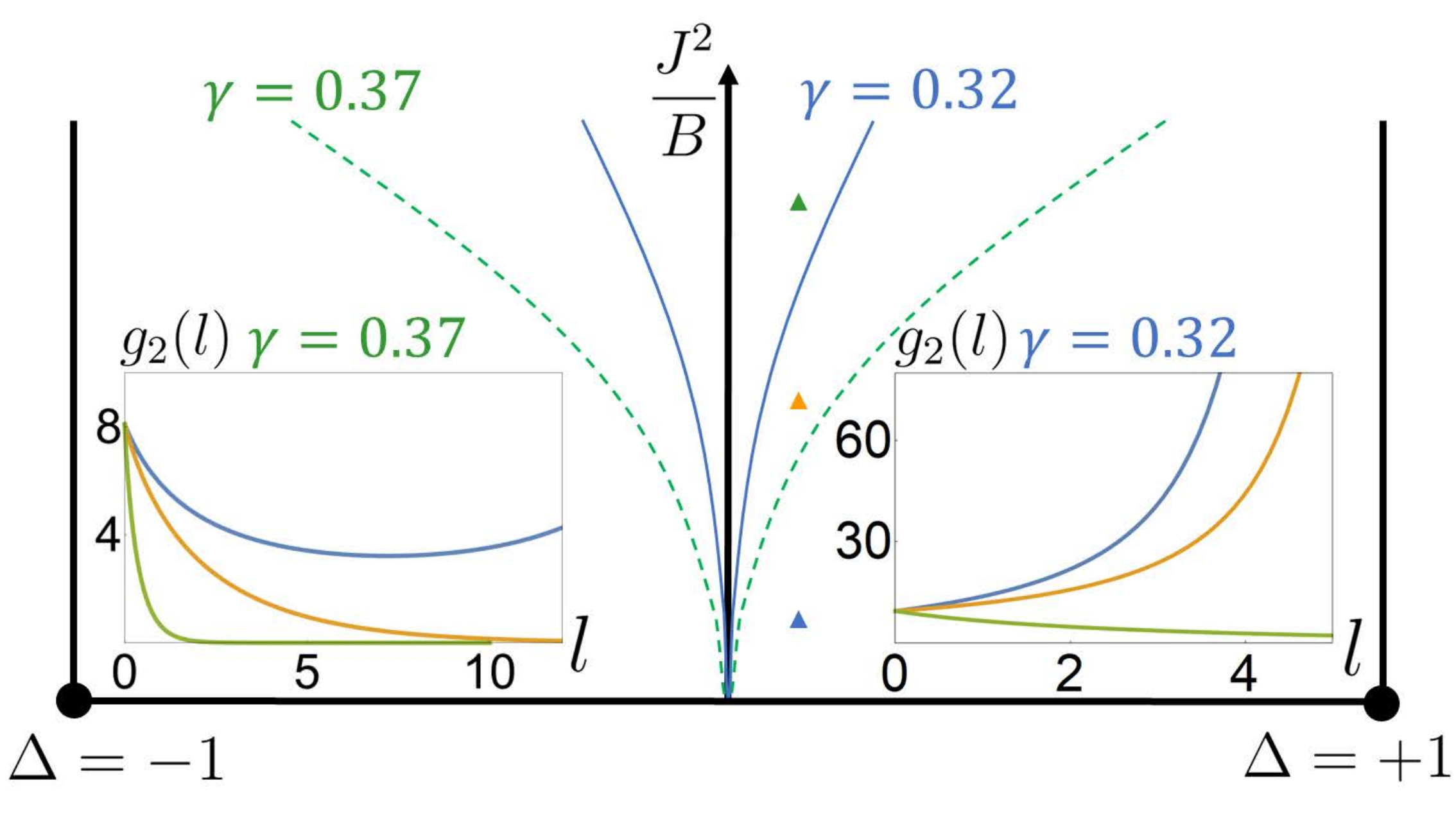}
%\\
%\subfloat[]{\includegraphics[width=0.45\textwidth]{paper writing/measurement phases section/general case EE scaling.pdf}\label{subfig:general J>0 gamma>0 numerics}}
\caption{Schematic phase diagram obtained from the 2-replica approximation RG flow (\ref{eq:SG RG flow 2}). A critical unitary strength $J_c$ separating the gapped area-law scaling phase ($J^2<J^2_c$) from the critical logarithmic phase ($J^2>J^2_c$) as reported for $\gamma/B=0.32$ (green dotted line) and  $\gamma/B=0.37$ (blue line). 
The left and right insets show the flow of $g_2$ under RG (\ref{eq:SG RG flow 2}) for $\gamma/B=0.37$ and $\gamma/B=0.32$ respectively evaluated at $J^2/B=0.019$ (blue),$J^2/B=0.20$ (orange), and $J^2/B=0.38$ (green), with $\Delta=0.07$ in all cases. 
The irrelevance of $g_2$ indicates a critical logarithmic scaling entanglement. 
%(b): average entanglement entropy $\overline{S}_0$  from numerical simulations to demonstrate the area law and logarithmic entanglement scaling phases for non-zero dimerization. The three red (blue) lines from bottom to top corresponds to increasing values of $J^2/B$. The values of $J^2/B$ was chosen to be (bottom to top) $J^2/B=0.019$, $J^2/B=0.20$ and $J^2/B=0.38$, and $\Delta=0.07$.
}
\label{fig:phases from numeric and 2 replica cal for general}
\end{figure}

For generic strong PPS case with all $|\Delta|>0$, $J^2>0$, and  $\gamma>0$, 
$g_2$ is the main parameter which controls the entanglement scaling. 
From a numerical solution of the RG flow Eq. \eqref{eq:SG RG flow root 8}, we see that for small initial values,  $g_2$,  flows to either irrelevant at large $J^2/B$ or grows indefinitely for sufficiently small values of $J^2/B$, (cf. Fig.~\ref{fig:phases from numeric and 2 replica cal for general}). 
In the latter case, since the term governed by $g_2$ is always more relevant than the $g_{\rho}$ and $g_{\phi}$ ones, the physics is entirely governed by the $g_2$ term, and the system is then in an area-law phase. 
When $g_2$ flows to zero at large $J^2/B$, the $g_{\epsilon}$ coupling term of $\cos(\sqrt{8}\phi)$ gaps at most one of the two sectors, leaving at least one sector gapless,  and the phase remains critical as discussed in Sec.~\ref{subsection: measurement only} for $\Delta=0$. 

The overall result for the phase diagram from the 2-replica approximation is schematically shown   in Fig.\ref{fig:phases from numeric and 2 replica cal for general}.
% from a numerical solution of the RG equations in \eqref{eq:SG RG flow 2}.
For a fixed partial post-selection  $\gamma/B \neq 0$, and non-zero dimerization $\Delta \neq 0$, we find critical values of $J^2$ beyond which $|g_2|$ grows indefinitely, corresponding to a critical-scaling phase. 
This phase expands when the retaining a larger subset   of quantum  trajectories (i.e. increasing  $\gamma/B$). The results from the RG analysis of the 2-replica model are also confirmed by the numerical evaluation of the entanglement entropy scaling in appendix~\ref{app:numerics}, Fig.~\ref{fig:phases from numerics for general}.
This expansion is understood as a result of the system exploring a larger extent of the Hilbert space as more trajectories are retained. This imposes fewer constraints on the unitary dynamics in generating large-scale entanglement, and
 is consistent with similar numerical findings with deterministic unitary dynamics~\cite{kells2023topological}.

\section{discussion and conlcusion}
\label{section:conclusions}

In this work, we have analysed the steady-state out-of-equilibrium phases of a monitored many-body quantum system when only part of the measurement readouts is retained (partial post-selection). 
We have first developed a general equation for the evolution of a quantum system under partial postselection of continuous Gaussian measurements, named Partial-Post-Selected Stochastic Schr\"odinger Equation (PPS-SSE) --- cf. eq.\eqref{eq:PPS SSE}, in which a parameter continuously bridges between the fully monitored and fully post-selected limits.
Since the two limits are known to give rise to MiPT of different universality classes, we have studied such crossover for a specific model of free Gaussian real fermions with random unitary dynamics. 
We analyzed the MiPT in a $2$-replica approximation which captures the simplest non-linearity in the system's state. 
Within the approximation, we derive the MiPT in terms of the low energy long-wavelength properties of and associated bosonised Hamiltonian in a $2$-replica-Choi-duplicated space in the limit of strong partial post-selection --- cf. Eq.\eqref{eq:bosonised Hamiltonian}.

We predict that in the strong PPS limit, the model presents MiPTs from area laws (with distinct quantum order) to a critical phase. 
% In the absence of unitary dynamics, the transition reduces to an Ising-like transition with a logarithmic critical scaling at the transition point.
We show that for strong yet finite partial post-selection, the phase diagram displays the same universal features as the post-selected model. In particular,
In the absence of unitary dynamics, the transition reduces to an Ising-like transition with a logarithmic critical scaling at the transition point. The entangling phase displays a $\log$ scaling 
instead of $\log^2$ in Ref.~\onlinecite{fava2023nonlinear}, with the only quantitative changes given by the expansion of the phase with critical scaling upon increasing the range of measurement outcomes retained --- cf. Fig.~\ref{fig:phases from numeric and 2 replica cal for general}.
Notably, our theory predictions are limited by the validity of the bosonization which breaks down at finite values of the partial post-selection indicating a possible phase transition at that point. 
Numerical results corroborate this finding by showing an abrupt change in the universal scaling of the measurement-only transition at a similar value of partial post-selection --- cf. Fig.~\ref{fig:critical exponent vary B}.

Our theory and its prediction shed new light on MiPT. 
First, the developed PPS-SSE is the first continuous stochastic equation that offers a novel analytical approach to study the relation between the critical phenomena observed in stochastic monitored dynamics and deterministic non-Hermitian evolution, as well as a means to analyze the transition between the two.
It can be therefore employed to explore the role of multiple trajectories in a variety of MiPTs.
The underlying microscopic derivation can also be the basis for obtaining similar PPS for other measurement-induced dynamics, like quantum jumps~\cite{turkeshi2021measurement,turkeshi2022entanglement,zerba2023measurement}.

Our findings for the Gaussian model indicate that the physics of post-selected measurement dynamics is robust against weak fluctuations induced by measurements. 
Our results suggest that different trajectories contribute different universal properties to the overall ensemble.
It is interesting to explore the generality of this finding and the mechanism underpinning the transition from post-selected to monitored dynamics identified in this work.

Finally, the feasibility of observing robust MiPTs by retaining a fraction of quantum trajectories provides a possible route to tackle the experimental post-selection problem, 
by performing tomography of the average state of a fraction of trajectories as opposed to tracking the trajectory-by-trajectory entanglement entropy.

\begin{acknowledgments}
We are grateful to M. Buchhold, S. Diehl, R. Fazio, M. Fava, A. Mesaros, M. Schiro, and K. Turkeshi for helpful discussions. A.R. acknowledges from the Royal Society, grant no. IECR2212041. 

\textit{Note added. During the completion of this manuscript, we became aware of a related work by Y. Le Gal, X. Turkeshi, and M. Schirò appearing on arxiv, numerically analyzing the stability of non-Hermitian dynamics in the context of quantum jump equation~\cite{Gal_to_appear}.}

\end{acknowledgments}

%Feng He for fruitful discussion on Bethe ansatz and extended Hubbard model

\appendix
\renewcommand{\thesection}{\Alph{section}}
\numberwithin{equation}{section}
\makeatletter
\section{PPS, shifted Gaussian and their time continuum limit}\label{sup:PPS and Gau}
\begin{figure}\label{fig: suppplement PPS section, rc, b and dependence on dt}
\subfloat[]{\includegraphics[width=0.45\textwidth]{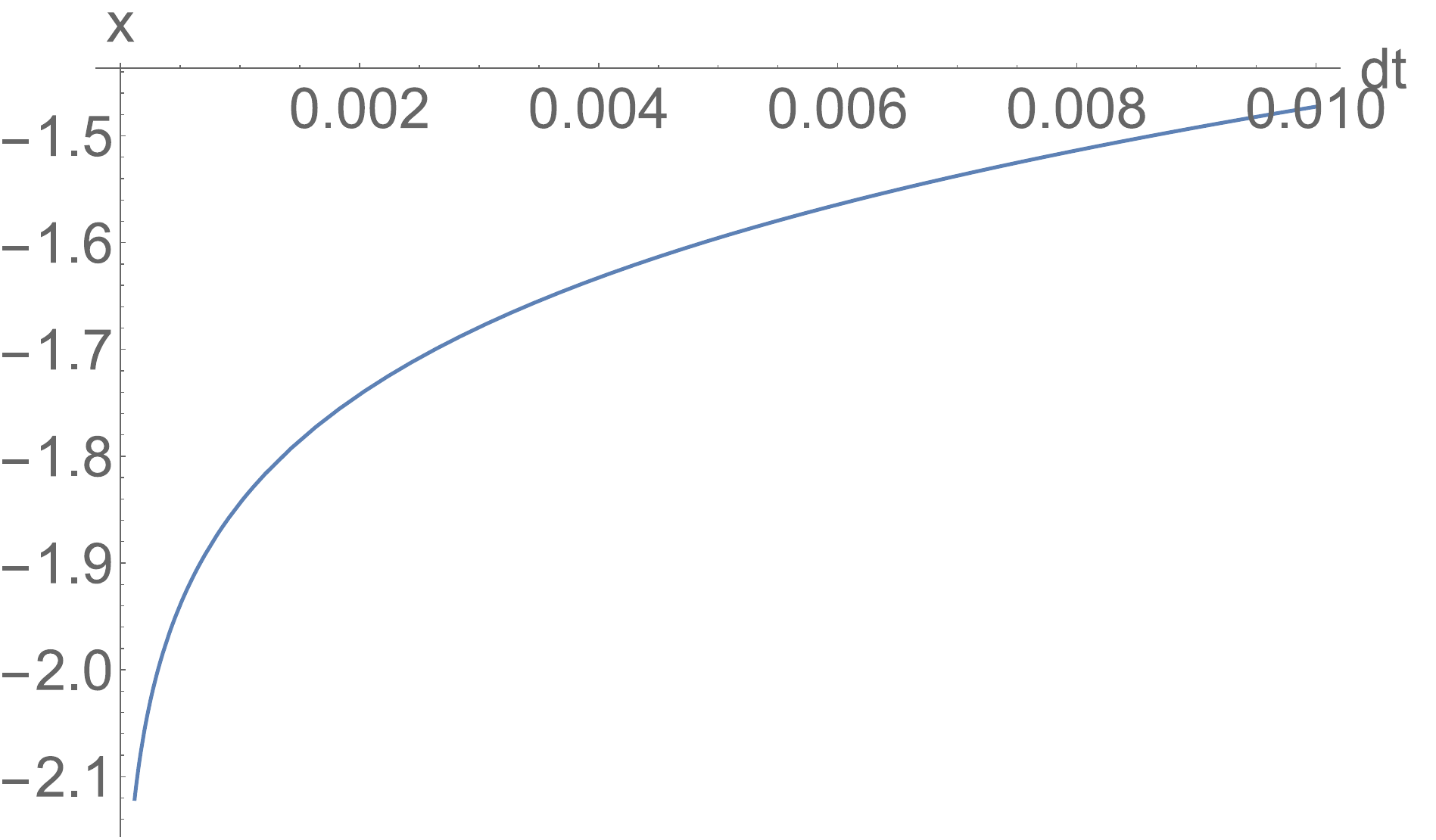}\label{subfig:rc dt depend}}\\
\subfloat[]{\includegraphics[width=0.45\textwidth]{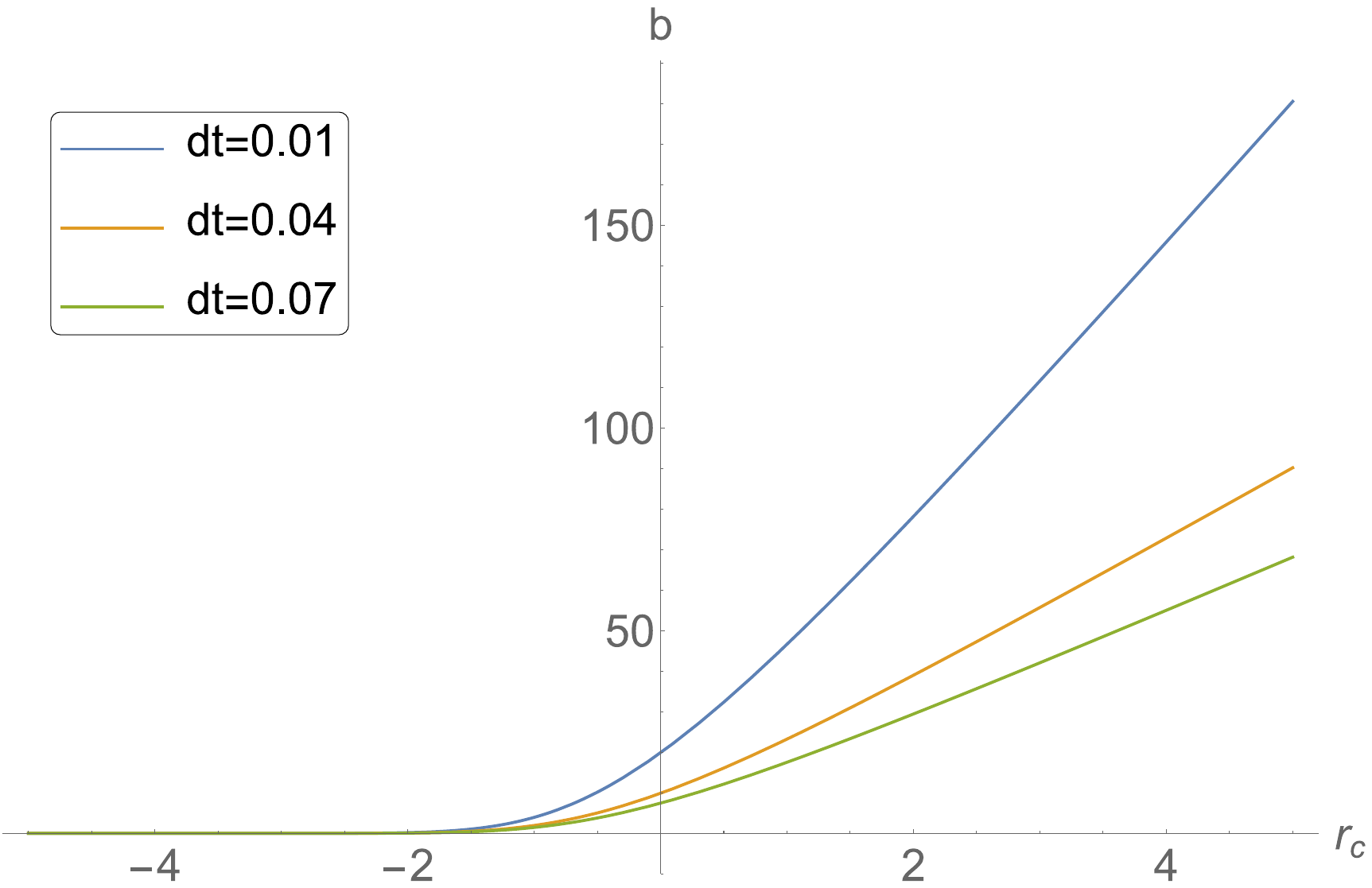}\label{subfig:b on rc}}
\caption{(a): dependence of $x$, from (\ref{eq:rc solve for dt}), on $dt$, with $b=1$ and $\langle\hat{O}_j\rangle=0.1$. It can be seen that $r_c\sim x$ approaches $-\infty$ in the time continuum limit $dt\rightarrow0$. (b): dependence of $b$, from (\ref{eq:rc solve for dt}), on $r_c$ for various $dt$. As $dt$ decreases, the same $b$ corresponds to a $r_c$ in the more negative direction.}
\end{figure}

Here, we demonstrate how the time continuum is taken giving $\delta\lambda=b\lambda$. From (\ref{eq:modified Stat}), the shift in mean $\delta\lambda$ of $P_{r_c}$ has the following $r_c$ dependence
\begin{eqnarray}\label{eq:mean_corr}
    \delta \lambda  =\Delta \sqrt{\frac{2}{\pi} }\frac{e^{-\frac{(-r_c+\lambda\langle \hat{O}_i\rangle)^2}{2\Delta^2}}}{1+{\rm Erf\left[\frac{-r_c+\lambda\langle \hat{O}_i\rangle}{\sqrt{2}\Delta}\right] }}
\end{eqnarray}
Since $\lambda$ scales as $\lambda\sim\sqrt{dt}$, we ask what is the $dt$ dependence we need to assign to $r_c $ in order that $\delta \lambda \sim \sqrt{dt} $ which matches the scaling of $\lambda$ in the Kraus operator. In other words, we are solving
\begin{eqnarray}\label{eq:rc solve for dt}
    \frac{e^{-(-x+a\langle\hat{O}_j\rangle\sqrt{dt})^2}}{1+{\rm Erf}\left[-x+a\langle\hat{O}_j\rangle\sqrt{dt} \right] }= ba\sqrt{dt} 
\end{eqnarray}
where $x = \frac{r_c(dt)}{\sqrt{2}\Delta} $, $a=\frac{\sqrt{\gamma}}{\sqrt{2}\Delta}$. This choice of parameterising $\delta\lambda$ ensures that $r_c$ depends negligibly on $\langle\hat{O}_j\rangle
$ as $dt\rightarrow0$ which is advantageous from an experimental point of view. The dependence of $r_c$ on $dt$ according to (\ref{eq:rc solve for dt}) is shown in figure(\ref{subfig:rc dt depend}) and it can be seen that $r_c\xrightarrow[]{dt\rightarrow0}-\infty$.
Analogous to $\lambda\xrightarrow[]{dt\rightarrow0}0$ and $\gamma$ is the parameter that captures measurement backaction, what captures PPS in time continuum is $b$ from (\ref{eq:rc solve for dt}) and its dependence on $r_c$ for fixed $dt$ is shown in figure(\ref{subfig:b on rc}), and it is lower bounded $b(r_c=-\infty)=0$. Solving for (\ref{eq:rc solve for dt}), we arrive at (\ref{eq:b definition}).

Under this scaling, we find that the correction to variance also scales like $\delta \sim \sqrt{dt} $. However unlike the unmodified mean $ \lambda $ which scales like $\lambda=\sqrt{\gamma dt } $, under this parametrization $ \Delta = {\cal O} (dt^0)$ and hence we can safely set $\delta \rightarrow 0 $.

\begin{figure}
\includegraphics[width=0.45\textwidth]{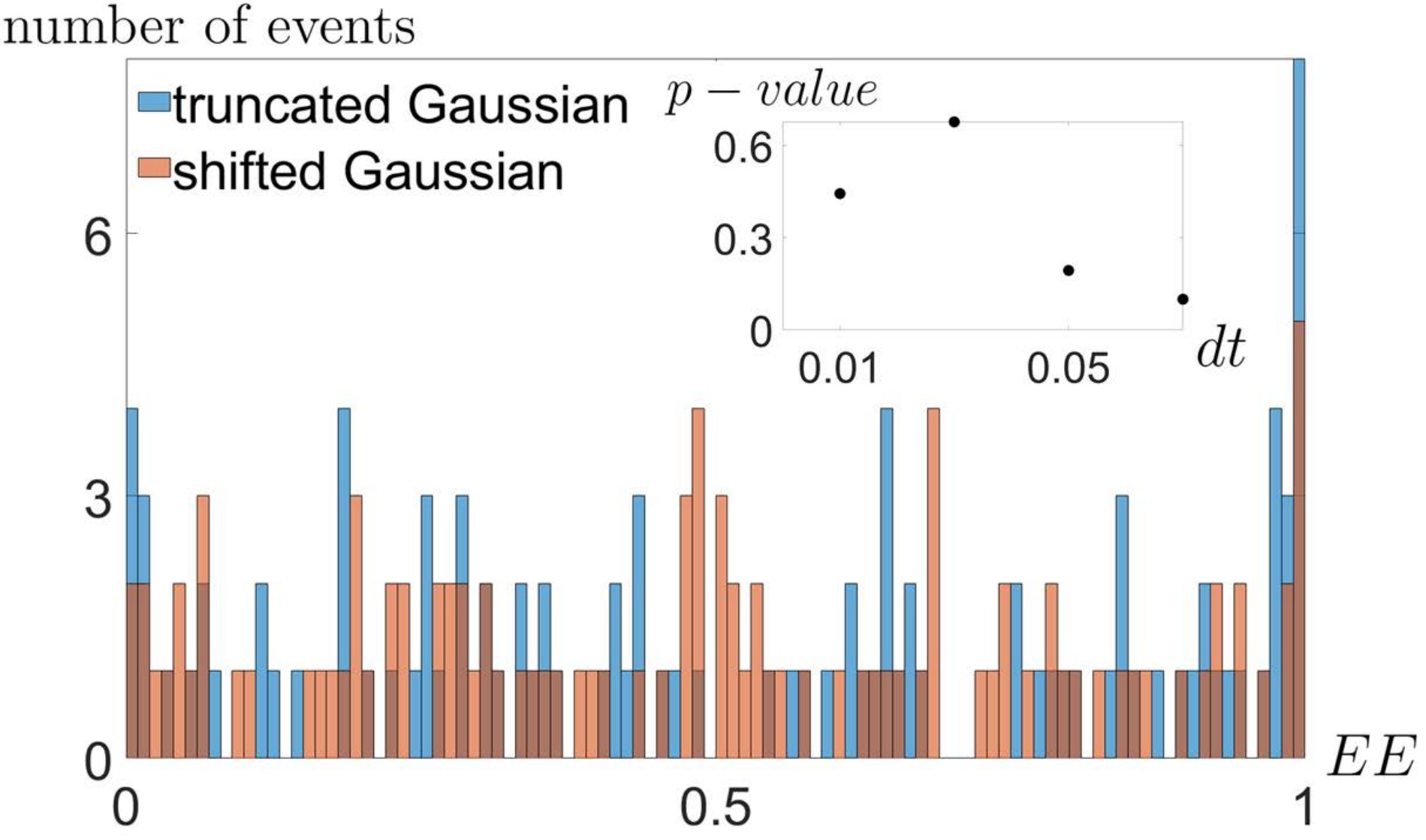}
\caption{Histogram of the steady state entanglement entropy distribution for various value of time increment in the numerics $\d t$. Blue colour are data evolved using truncated Gaussian distribution at $r_c$, red colour uses (\ref{eq:PPS SSE}). The parameters used for this histogram are $b=0.2, \gamma=0.5$ and $dt=0.01$. The inset shows $p$-values calculated for various $dt$ using the Two-sample Kolmogorov-Smirnov test, revealing an upward trend for decreasing $dt$ implying more overlapping between the data. For the values of $b$ and $dt$ considered in the histogram, the null hypothesis cannot be rejected and the 2 different sets of data are statistically indistinguishable.}
\label{fig:KS2 test histogram}
\end{figure}

In addition to the two-sample Kolmogorov-Smirnov test on the probability distribution in the main text, we have also verified numerically the shifted Gaussian approximation by considering a 2-qubit toy model. The toy model is described by the Hamiltonian $H=\sigma^+_1\sigma^-_2+\sigma^+_2\sigma^-_1$, and the 2 qubits are subject to measurement operators $(\mathbb{I}+(-1)^j\sigma_j^z)/2,j=1,2$. Firstly for fixed $b$ in (\ref{eq:b definition}), two separate distributions of the steady state entanglement entropy is computed via 2 different ways: 1. the update of the state by the measurement operators is given by (\ref{eq:Kraus for continuous}) with the probability distribution $P_{r_c}(x_j)$ given by the truncated Gaussian, 2. the update of the state is computed via (\ref{eq:PPS SSE}). Then, the 2 distribution are compared using the Two-sample Kolmogorov-Smirnov test. This is repeated for different values of $\Delta t$, the time increment used. The results are shown in figure (\ref{fig:KS2 test histogram}).

\begin{figure}
\subfloat[]{\includegraphics[width=0.45\textwidth]{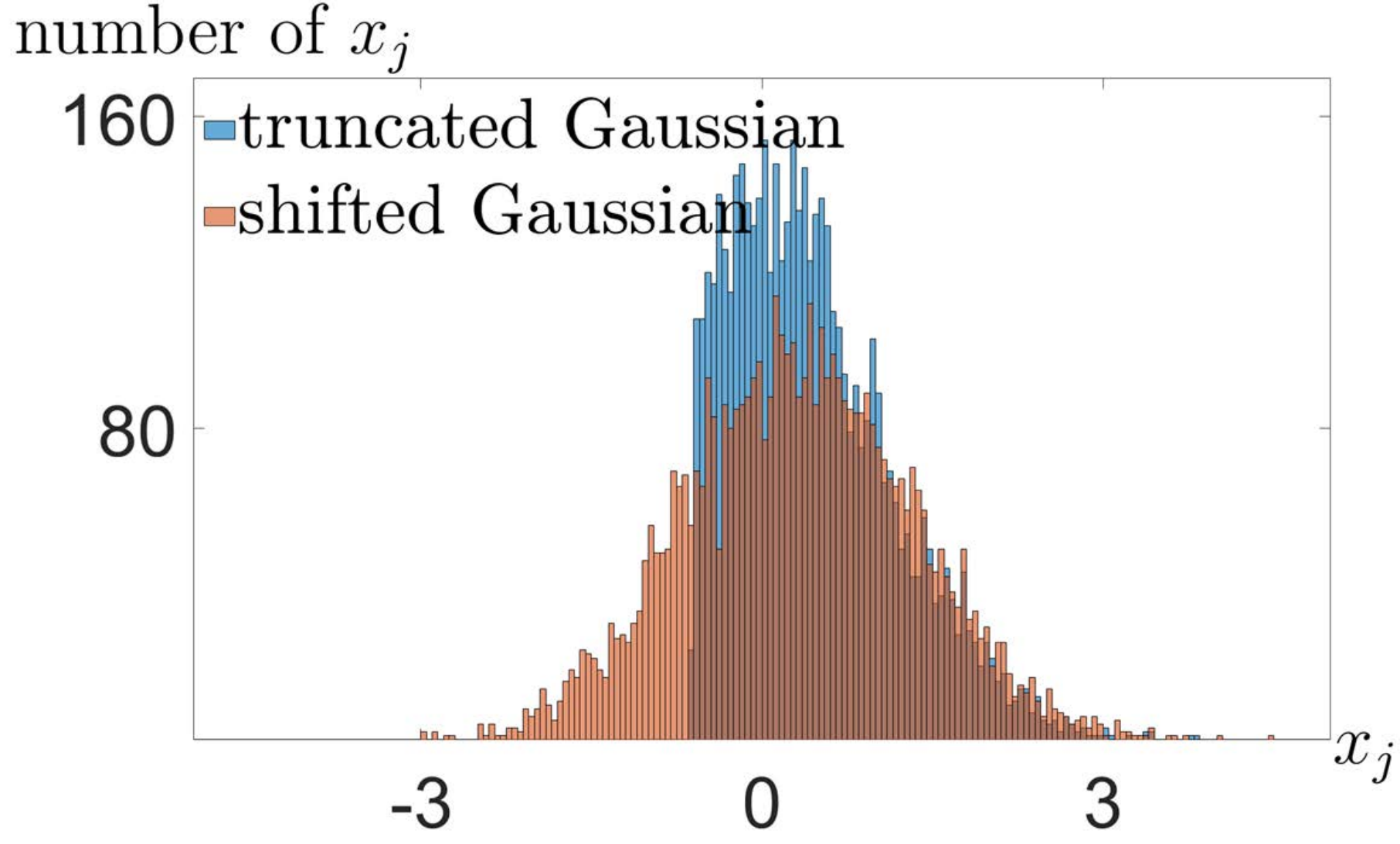}\label{subfig:prob_dist_Gaussian_dt_0.05}}\\
\subfloat[]{\includegraphics[width=0.45\textwidth]{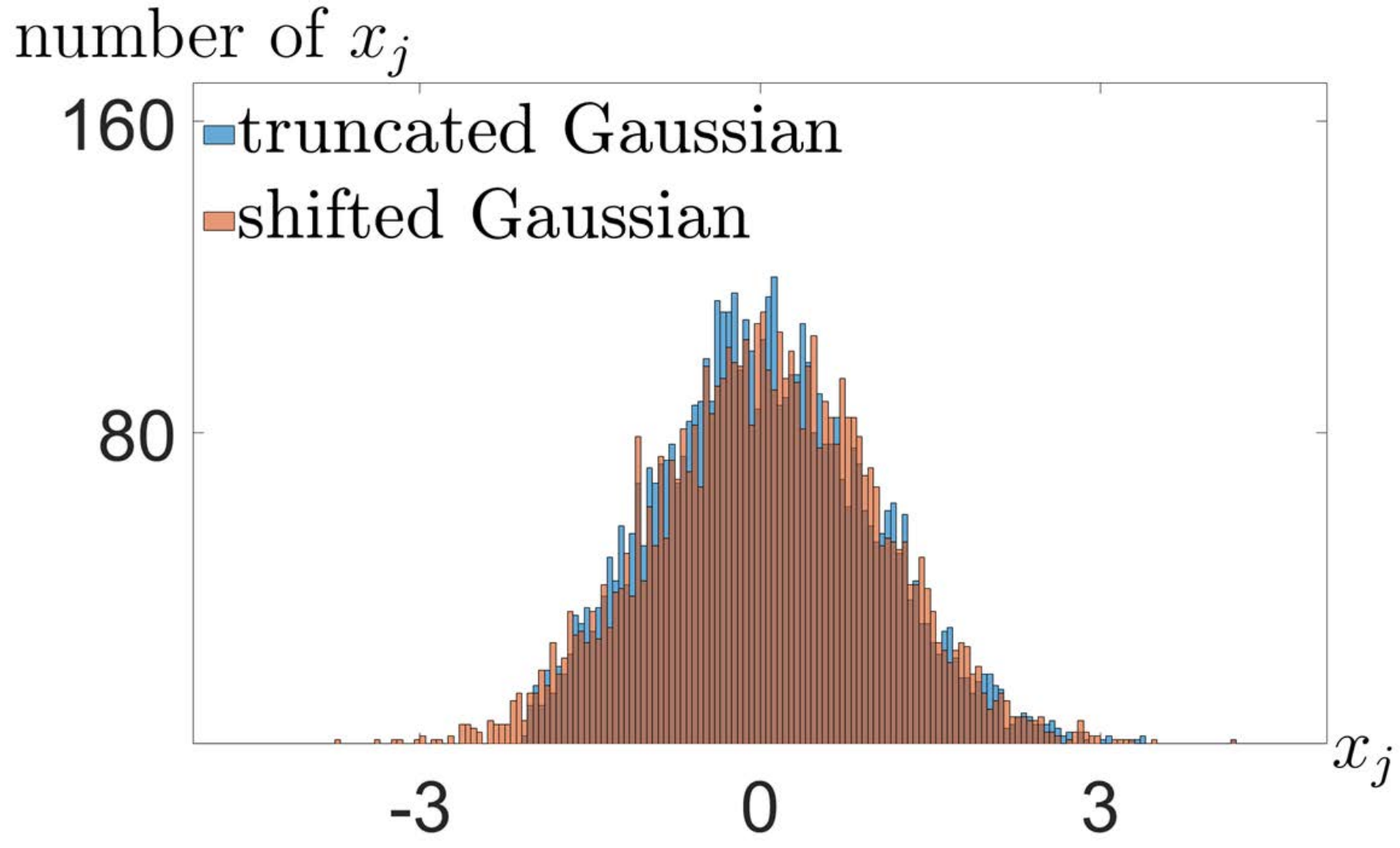}\label{subfig:prob_dist_Gaussian_dt_0.001}}
\caption{Histograms of the truncated Gaussian $P_{r_c}(x_j)$ and shifted Gaussian $\underline{P}(x_j)$ in (\ref{eq:modified Stat}). (a): samplings drawn from the truncated Gaussian (blue) and the shifted Gaussian (red) for $dt=0.05$ and the null hypothesis is rejected at a significance level of 0.05. The p-value from KS2 sample test is $0.00$. (b): similar to (a) but samples generated with $dt=0.001$. The p-value is $0.31$ and the null hypothesis cannot be rejected, indicating the distribution is statistically indistinguishable. The values of the parameters used here are $\langle O_j\rangle=1,b=1$ and $\gamma=0.5$ and we used $5000$ samplings of the distributions.}
\label{fig:histogram for 2 r_c and shifted prob dist}
\end{figure}

For completeness, we also display numerically the samplings from the truncated and shifted Gaussian in figure (\ref{fig:histogram for 2 r_c and shifted prob dist}), together with the associated p-values calculated. It clearly display statistical equivalence for $dt=0.001$.

\section{Continuous measurement as non-Herm noises and PPS on Gaussian average}\label{sup:cont mea as nH and PPS}

The procedures we are using here is an extension to~\cite{fava2023nonlinear}, and we make explicit link to the discrete time description of continuous measurement in (\ref{eq:Kraus for continuous}), and extending it to PPS. To begin with, we start from (\ref{eq:Kraus for continuous}) and changes some of the factor slightly for later convenience:
\begin{align}\label{eq:supp cont mea, Kraus}
    \hat{k}_j(x,\lambda)&=\mathcal{N}_j\exp\left(-\frac{(x-2\lambda\hat{O}_j)^2}{4\Delta^2}\right)\nonumber \\\hat{k}_j(x,\lambda)\ket{\psi_t}&=\tilde{\mathcal{N}}_j\exp\left(-\frac{x^2}{4\Delta^2}\right)\exp\left(\frac{\lambda\hat{O}_jx}{\Delta^2}\right)\ket{\psi_t}
\end{align}
An initial normalised density matrix $\rho_0$ is updated as
\begin{align}
    \frac{\hat{k}_j(x,\lambda)\rho_0\hat{k}^{\dagger}_j(x,\lambda)}{\Tr[\hat{k}_j(x,\lambda)\rho_0\hat{k}^{\dagger}_j(x,\lambda)]}=\frac{\check{\rho}_{x,\lambda}}{\Tr[\check{\rho}_{x,\lambda}]}
\end{align}
The average density matrix $\overline{\rho}$ across all measurement outcome at a particular time step is
\begin{align}\label{eq:supp cont mea, ave rho from contin mea rho}
    \overline{\rho}&=\int_{-\infty}^{\infty} dx \frac{\check{\rho}_{x,\lambda}}{\Tr[\check{\rho}_{x,\lambda}]}P(x,\lambda) \nonumber \\
    \overline{\rho}&=\int_{-\infty}^{\infty} dx \hat{k}_j(x,\lambda)\rho_0\hat{k}^{\dagger}_j(x,\lambda) \nonumber \\
    &=\tilde{\mathcal{N}}_j^2 \int_{-\infty}^{\infty} dx \exp\left(-\frac{x^2}{2\Delta^2}\right)\exp\left(\frac{x\lambda\hat{O}_j}{\Delta^2}\right)\rho_0\exp\left(\frac{x\lambda\hat{O}_j}{\Delta^2}\right)
\end{align}
which implies
\begin{align}\label{eq:supp cont mea, Kraus norm}
    \int_{-\infty}^{\infty} dx \hat{k}^{\dagger}_j(x,\lambda)\hat{k}_j(x,\lambda)&=\mathbb{I} \nonumber \\
    \tilde{\mathcal{N}}_j^2 \int_{-\infty}^{\infty} dx \exp\left(-\frac{x^2}{2\Delta^2}\right)\exp\left(\frac{x2\lambda\hat{O}_j}{\Delta^2}\right)&=\mathbb{I}
\end{align}
Rewriting 
\begin{align}\label{eq:supp cont mea, rename var}
    \Delta^2=\Delta'^2\lambda/\delta t,x=M_j\Delta'^2 \text{and} \ \gamma=\lambda/\Delta'^2=\lambda^2/\Delta^2\delta t
\end{align}
(\ref{eq:supp cont mea, Kraus norm}) becomes
\begin{align}
    \tilde{\mathcal{N}}_j^2 \int_{-\infty}^{\infty} dM_j \exp\left(-\frac{M_j^2\delta t}{2\gamma}\right)\exp\left(2M_j\hat{O}_j\delta t\right)&=\mathbb{I}
\end{align}
and one can interpret that the Kraus operator is alternatively described by
\begin{align}\label{eq:supp con mea, interpreted as nh noise}
    \overline{\hat{k}}(M_j)=\exp\left(M_j\hat{O}_j\delta t\right) \nonumber \\
    \int_{-\infty}^{\infty}d\mu(M_j)\overline{\hat{k}}^{\dagger}(M_j)\overline{\hat{k}}(M_j)=\mathbb{I}
\end{align}
and normalised over the Gaussian measure $d\mu(M_j)\propto dM_j \exp\left(-\frac{M_j^2\delta t}{2\gamma}\right)$, which describes a Gaussian random variable $M_j$ with mean $\mathbb{E}_G[M_j]=0$ and variance $\mathbb{E}_G[M_j^2]=\gamma/\delta t$.

The readout of a continuous measurement is now represented by the variable $M_j$, and its backaction on the system is given by the Kraus operator in \eqref{eq:supp con mea, interpreted as nh noise}. To generalise it to a time process,  we first give $M_j(t_l)$ a time index $t_l=l\delta t$. Then, an initial density matrix $\rho_0$ evolves from time $t_0=0$ to $t_N=T=N\delta t$ as 
\begin{align}\label{eq:supp con mea, nH Gaussian}
    \check{\rho}_{M}(T)=\prod_{l=1}^{l=N}\overline{\hat{k}}(M_j(t_l))\rho(0)\overline{\hat{k}}^{\dagger}(M_j(t_l)),
\end{align}
where $M$ labels the quantum trajectory. In the time continuum limit, \eqref{eq:supp con mea, nH Gaussian} becomes
\begin{align}\label{eq:supp con mea, rand nH Ham demo}
    &\check{\rho}_{M}=K(t)\rho(0)K^{\dagger}(t) \nonumber \\
    &K(t)=\exp[-i\int_{0}^{T}dt'H(t')]=\exp[-i\int_{0}^{T}dt'i M_j(t')\hat{O}_j]
\end{align}
and $\mathbb{E}_G[M_j(t)]=0, \,\mathbb{E}_G[M_j(t)M_j(t')]=\gamma\delta(t-t')$. From \eqref{eq:supp con mea, rand nH Ham demo}, we observe that the overall effect of a continuous measurement generates a random non-Hermitian Hamiltonian $H(t)=iM_j(t)\hat{O}_j$ in time. Generalisation to multiple measurement $j=1\dots L$, and including another competing set, follows the same line as each process is independent to each other, and we arrive at \eqref{eq:OG majorana model} in the main text.

In the case of PPS, we saw, in Appendix \ref{sup:PPS and Gau}, that PPS shifts the mean of the random variable $x$ by $b\lambda$. We can interpret this as a shift in the mean of the measure. Using the relationship (\ref{eq:supp cont mea, rename var})
\begin{align}
    d\mu(M_j)&\xrightarrow[]{PPS}dx \exp\left(-\frac{(x-b\lambda)^2}{2\Delta^2}\right)\nonumber \\ &\propto dM_j \exp\left(-\frac{(M_j-b\gamma)^2\delta t}{2\gamma}\right)
\end{align}
we arrive at the final line where the mean of the Gaussian average is shifted $\mathbb{E}_G[M_j]=b\gamma=B$, as required from PPS. The generalisation to multiple weak continuous measurement is straightforward, with different measurement corresponds to different non-Hermitian noises and taking the time continuum limit gives (\ref{eq:Kraus, dm evolve}) (since noises from different measurements are independent from each other, cross product between different noise vanishes in time continuum limit).

\section{operator-state correspondence and replica majorana Hamiltonian}\label{sup:Choi and replica trick}
In this appendix and below, we distinguish the ket or bra space in the Choi–Jamiołkowski isomorphism by $\sigma=\pm$ to replace $\uparrow$ and $\downarrow$ in the main text.

The Choi–Jamiołkowski isomorphism maps an operator to a duplicated Hilbert space:
\begin{align}\label{eq:Choi display}
    \hat{O}=\sum_{i,j}O_{i,j}\ket{i}\bra{j}\xrightarrow[]{Choi}\sum_{i,j}O_{i,j}\ket{i}\otimes\ket{j}=\ket{\hat{O}}
\end{align}

Under Choi–Jamiołkowski isomorphism, the trace operation between 2 operators becomes a transition amplitude:
\begin{align}\label{eq:trace in Choi}
    \text{Tr}[\hat{A}^{\dagger}\hat{B}]\xrightarrow[]{Choi}\langle\hat{A}|\hat{B}\rangle
\end{align}
hence equ (\ref{eq:Choi ave as trans}). When dealing with density matrix $\rho$, the action of some operator on the density matrix becomes action on the Choi state:
\begin{align}\label{eq:operator in Choi}
    \hat{A}\rho&=\sum_{i,j,k}A_{i,j}\rho_{j,k}\ket{i}\bra{k}\xrightarrow[]{Choi}\sum_{i,j,k}A_{i,j}\rho_{j,k}\ket{i}\ket{k}=\hat{A}\otimes\mathbb{I}\ket{\rho} \nonumber \\
    \rho\hat{A}&\xrightarrow[]{Choi}\mathbb{I}\otimes\hat{A}^{T}\ket{\rho}, \ \hat{B}\rho\hat{A}\xrightarrow[]{Choi}\hat{B}\otimes\hat{A}^{T}\ket{\rho}
\end{align}
and hence (\ref{eq:Choi Kraus}). Writing out explicitly the average dynamics of $n$-replica described by (\ref{eq:Choi n replica}):
\begin{align}\label{eq:Majorana before Klein}
    &\mathbb{E}^{(PPS)}_G[\left(K(t)\otimes K^*(t)\right)^{\otimes n}]\ket{\rho^n(0)}\rangle\nonumber \\ =&\mathbb{E}^{(PPS)}_G\left[\exp\left(-i\int_0^tH_n(t')dt'\right)\right] \nonumber \\ H_n(t')&=\sum_{\substack{\sigma=\pm\\a=1\dots n}}\sum_j\left[J_j(t')+i\sigma M_{j}(t')\right]i\chi_j^{(\sigma a)}\chi_{j+1}^{(\sigma a)}
\end{align}
where $\sigma$ distinguish ket/bra space, $a$ for replica: $\chi_j^{(-a)}=\mathbb{I}^{\otimes 2a+1}\otimes\chi_j^*\otimes\mathbb{I}^{\otimes 2a}, \chi_j^{(+a)}=\mathbb{I}^{\otimes 2a}\otimes\chi_j\otimes\mathbb{I}^{\otimes 2a+1}$.
Up to this point, the Majorana operator $\chi_j^{(\sigma a)}$ is not well defined as they anti-commute within the same branch and replica, while commuting each other in different branch or replica. To resolve this, one should first map the Majorana to spin Hilbert space, then define new Majorana operators which differs from the one in (\ref{eq:Majorana before Klein}) by a Klein factor which plays the role of Pauli string in the replica space~\cite{fava2023nonlinear}.
Calling the proper Majorana operators in the replica space as $\chi_j'^{(\sigma a)}, \ \{\chi_j'^{(\sigma a)},\chi_l'^{(\sigma' a')}\}=\delta_{j,l}\delta_{\sigma,\sigma'}\delta_{a,a'}$, the bilinear product $\chi_j^{(\sigma a)}\chi_{j+1}^{(\sigma a)}=\chi_j'^{(\sigma a)}\chi_{j+1}'^{(\sigma a)}$ is preserved with no extra factor.

To evaluate (\ref{eq:Majorana before Klein}), we expand it using cumulant expansion up to 2nd order $\langle e^A\rangle\approx\exp[\langle A\rangle+1/2(\langle A^2\rangle-\langle A\rangle^2)]$ and note that the Guassian measure now centred at $b\gamma_j$. With a slight abuse of notation by calling the new anti-commuting Majorana as $\chi_j'^{(\sigma a)}\rightarrow\chi_j^{(\sigma a)}$, for $n=2$ we arrive at (\ref{eq:2 replica effective majorana Hamiltonian}).

Finally, the boundary state $\ket{\mathcal{C}_{2,\mathbf{A}}}\rangle,\ket{\mathbb{I}}\rangle$ have the following properties with the Pauli matrices in the replica space $\sigma_{\alpha,j}^{(a)}\ ,\alpha=x,y,z \ a=1,2$~\cite{fava2023nonlinear}:
\begin{widetext}
\begin{align}\label{eq:bound state proporties}
    \sigma_{\alpha,j}^{(a)}\mathbb{I}\sigma_{\alpha,j}^{(a)}=\mathbb{I}&\xrightarrow[]{Choi}i\chi_j^{(+a)}\chi_j^{(-a)}\ket{\mathbb{I}}\rangle=\ket{\mathbb{I}}\rangle \nonumber \\
    \sigma_{\alpha,j}^{(a)}\mathcal{C}_{2,\mathbf{A}}\sigma_{\alpha,j}^{(a)}=\mathcal{C}_{2,\mathbf{A}}&\xrightarrow[]{Choi}i\chi_j^{(+a)}\chi_j^{(-a)}\ket{\mathcal{C}_{2,\mathbf{A}}}\rangle=\ket{\mathcal{C}_{2,\mathbf{A}}}\rangle ,\ j \notin \mathbf{A} \nonumber \\
    \sigma_{\alpha,j}^{(2)}\mathcal{C}_{2,\mathbf{A}}\sigma_{\alpha,j}^{(1)}=\mathcal{C}_{2,\mathbf{A}}&\xrightarrow[]{Choi}-i\chi_j^{(+2)}\chi_j^{(-1)}\ket{\mathcal{C}_{2,\mathbf{A}}}\rangle=\ket{\mathcal{C}_{2,\mathbf{A}}}\rangle ,\ j \in \mathbf{A}\nonumber \\
    \sigma_{\alpha,j}^{(1)}\mathcal{C}_{2,\mathbf{A}}\sigma_{\alpha,j}^{(2)}=\mathcal{C}_{2,\mathbf{A}}&\xrightarrow[]{Choi}i\chi_j^{(+1)}\chi_j^{(-2)}\ket{\mathcal{C}_{2,\mathbf{A}}}\rangle=\ket{\mathcal{C}_{2,\mathbf{A}}}\rangle ,\ j \in \mathbf{A}
\end{align}
\end{widetext}

\section{Solution for 2-replica for monitored case}\label{sup:2 replica monitor}
Recall that the effective Hamiltonian without PPS reads

\begin{align}\label{eq:2 replica Ham no PPS}
    \mathcal{H}&=\frac{1}{2}\sum_j J^2(\sum_{\substack{\sigma=\pm\\a=1,2}}i\chi^{(\sigma a)}_j\chi^{(\sigma a)}_{j+1})^2 - \gamma_j(\sum_{\substack{\sigma=\pm\\a=1,2}}\sigma i\chi^{(\sigma a)}_j\chi^{(\sigma a)}_{j+1})^2
\end{align}
one can write it entirely as local SO(4) generators defined using Majorana operators:
\begin{align}\label{eq:SO(4) gen maj}
    S^{\alpha,\beta}_j=\frac{i}{2}\comm{\chi_j^{\alpha}}{\chi_j^{\beta}}
\end{align}
and for generic $J^2,\gamma$, only a subset of (\ref{eq:SO(4) gen maj}) commutes with $\mathcal{H}$~\cite{fava2023nonlinear,bao2021symmetry}. An important set of local symmetry which will become clear are the local on-site parity $\mathcal{R}_j=\prod_a i\gamma^{(+a)}\gamma^{(-a)},\comm{\mathcal{R}_j}{\mathcal{H}}=0$.

One can readily define the following spin-1/2 operators:
\begin{align}\label{eq:Sigma spin}
    \Sigma^{\mu}=\frac{1}{2}\underline{\mathbf{c}}_j^{\dagger}\sigma_{\mu}\underline{\mathbf{c}}_j
\end{align}
where $\sigma_{\alpha}\ ,\alpha=x,y,z$ are the usual Pauli matrices and $\underline{\mathbf{c}}_j=(c_{j,\uparrow},c_{j,\downarrow})^T$. The other spin-1/2 generators is associated with the $\eta
$ spin in the Hubbard model, generated via Shiba transformation~\cite{yang1990so,essler2005one}.
\begin{align}\label{eq:eta spin}
    \eta_j^z=\frac{1}{2}\left(c^{\dagger}_{j,\uparrow}c_{j,\uparrow}+c^{\dagger}_{j,\downarrow}c_{j,\downarrow}-1\right) \ , \ \eta_j^+=c^{\dagger}_{j,\uparrow}c^{\dagger}_{j,\downarrow}
\end{align}
these two species of SU(2) generators stems from the fact that SO(4)$\cong[$SU(2)$\times$SU(2)$]/\mathbb{Z}_2$. The quotient by $\mathbb{Z}_2$ comes from the criterion that
\begin{align}\label{eq:SU2 SU2 quotient}
    \sum_j\left[\eta_j^z+\Sigma_j^z\right]=\sum_jc^{\dagger}_{j,\uparrow}c_{j,\uparrow}-\frac{L}{2} \ \in \mathbb{Z}
\end{align}
and $\eta_j^z,\Sigma_j^z$ can either be both integer or both half-integer (assuming $L$ is even). Recalling that the local parity operator $\mathcal{R}_j=\pm1$ commutes with $\mathcal{H}$, constructing the projector $\Pi_{j,+}=\frac{1}{2}\left(1+\mathcal{R}_j\right)$ we observe that
\begin{align}
    \Pi_{j,+}\Sigma_j^{\mu}\Pi_{j,+}=0 \ \text{, while} \ \Pi_{j,+}\eta_j^{\mu}\Pi_{j,+}=\eta_j^{\mu}
\end{align}
hence the two different SU(2) $\Sigma^{\mu},\eta^{\mu}$ acts on $\mathcal{R}_j=\mp1$ sector respectively. The choice of the initial state $\ket{\rho(0)}\rangle=\ket{\mathbb{I}}\rangle,\mathcal{R}_j\ket{\mathbb{I}}\rangle=+1\ket{\mathbb{I}}\rangle$ fixes the sector and should match the sector the boundary state is in. To complete the prove that $\eta^{\mu}$ (and hence $\Sigma^{\mu}$) are spin-1/2 operators, we demonstrate that the total spin operator has eigenvalue:
\begin{align}
    {\eta_j^x}^2+{\eta_j^y}^2+{\eta_j^x}^2\ket{\mathbb{I}}\rangle=\frac{4}{3}\ket{\mathbb{I}}\rangle=S(1+S)\ket{\mathbb{I}}\rangle
\end{align}
where $S=1/2$.

The SO(4) generators in (\ref{eq:SO(4) gen maj}) can be expressed in term of these two SU(2) generators i.e. $S^{+1,+2}=2(\Sigma^z+\eta^z)$. Writing (\ref{eq:2 replica Ham no PPS}) in terms of (\ref{eq:eta spin}) and (\ref{eq:Sigma spin}), we arrive at \eqref{eq:2 XXZ Ham}
and the physics can readily be extracted via usual means i.e. Bethe Ansatz and bosonisation.

\section{effective spin Hamiltonian and bosonisation details}\label{sup:Bosonisation}
As mentioned in the main text, there are two conserved charges $\comm{\sum_j\gamma_j^{(\sigma1)}\gamma_j^{(\sigma2)}}{\mathcal{H}}$, which suggest the following 2 complex fermions:
\begin{align}
c^{\dagger}_{j,\uparrow}=\frac{\gamma_j^{(+1)}+i\gamma_j^{(+2)}}{2} \ , \ c^{\dagger}_{j,\downarrow}=\frac{\gamma_j^{(-1)}-i\gamma_j^{(-2)}}{2}
    \label{complex fermions}
\end{align}
Written in terms of the complex fermions, it becomes $\comm{\mathcal{H}}{N_{\sigma}}=0, \ N_{\sigma}=\sum_j c^{\dagger}_{j,\sigma}c_{j,\sigma}, \ \sigma=\uparrow ,\downarrow $. These two conserved $U(1)$ charges will be the basis for abelian bosonisation later. Inserting this relationship and followed by an unitary transformation $c^{\dagger}_{j,\uparrow}\rightarrow(i)^jc^{\dagger}_{j,\uparrow}, \ c^{\dagger}_{j,\downarrow}\rightarrow(-i)^jc^{\dagger}_{j,\downarrow}$, the Majorana operators are transformed as:
\begin{align}
    -i(\gamma^{(+1)}_j\gamma^{(+1)}_{j+1}+\gamma^{(+2)}_j\gamma^{(+2)}_{j+1})\rightarrow -2(c^{\dagger}_{j,\uparrow}c_{j+1,\uparrow}+c^{\dagger}_{j+1,\uparrow}c_{j,\uparrow}) \nonumber \\
    i(\gamma^{(-1)}_j\gamma^{(-1)}_{j+1}+\gamma^{(-2)}_j\gamma^{(-2)}_{j+1})\rightarrow -2(c^{\dagger}_{j,\downarrow}c_{j+1,\downarrow}+c^{\dagger}_{j+1,\downarrow}c_{j,\downarrow})
    \label{overall c change}
\end{align}

Inserting these into (\ref{eq:2 replica effective majorana Hamiltonian}), we arrive at (\ref{eq:effective spinful fermion Hamiltonian}).

We now proceed to bosonise (\ref{eq:effective spinful fermion Hamiltonian}) w.r.t. the basis $\sigma=\uparrow,\downarrow$. We first compute terms corresponding to no dimerisation $\mathcal{O}(\Delta)^0$. $H_0$ the kinetic part gives the usual free Luttinger liquid Hamiltonian with $K=1$:
\begin{align}
    H_0=\frac{v_F}{2\pi}\sum_{\sigma=\uparrow,\downarrow}\int_x(\partial_x\theta_{\sigma})^2+(\partial_x\phi_{\sigma})^2
\end{align}
With bosonisation, we can investigate the strong PPS limit where $J^2,\gamma\ll B$ . This is the limit at which the excitation is small compare to the Fermi energy and bosonisation remains valid. As bosonising a lattice model will inevitably generate term whose appearance depends directly on the filling fraction, the filling fraction is determined by utilising the properties in (\ref{eq:bound state proporties}) which gives
\begin{align}\label{eq:half filling show}
    \langle\bra{\mathbb{I}}i\chi_j^{(+1)}\chi_j^{(-2)}\ket{\mathbb{I}}\rangle=0 \nonumber \\
    \langle\bra{\mathbb{I}}-\chi_j^{(+1)}\chi_j^{(+2)}\ket{\mathbb{I}}\rangle=0 \nonumber \\
    \langle\bra{\mathbb{I}}(c^{\dagger}_{j,\uparrow}+c_{j,\uparrow})(c^{\dagger}_{j,\uparrow}-c_{j,\uparrow})\ket{\mathbb{I}}\rangle=0\nonumber \\
    \langle\bra{\mathbb{I}}1-2c^{\dagger}_{j,\uparrow}c_{j,\uparrow})\ket{\mathbb{I}}\rangle=0
\end{align}
similarly for $c^{\dagger}_{j,\downarrow}$ and the boundary state $\ket{\mathcal{C}_{2,\mathbf{A}}}\rangle$. Therefore, this specifies that we are dealing with half filling $k_F=\pi/2$ and some term that oscillates with $e^{4ik_Fx}$ should in fact be kept.

The term $H_{umk}$ in (\ref{eq:effective spinful fermion Hamiltonian}) becomes
\begin{align}
    H_{umk}\propto&\sum_{\substack{\sigma=\uparrow,\downarrow\\ j}}(c^{\dagger}_{j,\sigma}c_{j+1,\sigma}+c^{\dagger}_{j+1,\sigma}c_{j,\sigma})^2 \nonumber \\
    =&-2\sum_{\substack{\sigma=\uparrow,\downarrow\\ j}}(c^{\dagger}_{j,\sigma}c_{j,\sigma}-\frac{1}{2})(c^{\dagger}_{j+1,\sigma}c_{j+1,\sigma}-\frac{1}{2})
    \nonumber \\
    \approx&-2a\sum_{\sigma=\uparrow,\downarrow}\int_x \frac{2}{\pi^2}(\partial_x\phi_{\sigma})^2 - \frac{2}{(2\pi\alpha)^2}cos4\phi_{\sigma}
\label{eq:H Umklapp boson}
\end{align}
while $H_m$ gives
\begin{align}\label{eq:Hm boson}
    H_m&\propto\sum_j (c^{\dagger}_{j,\uparrow}c_{j+1,\uparrow}+h.c.)(c^{\dagger}_{j,\downarrow}c_{j+1,\downarrow}+h.c.) \nonumber \\
    \approx&a\int_x\Bigg[\frac{4}{2\pi}\nabla\phi_{\uparrow}+\frac{e^{2ik_Fx}}{2\pi\alpha}2ie^{-i2\phi_{\uparrow}(x)}-\frac{e^{-2ik_Fx}}{2\pi\alpha}2ie^{i2\phi_{\uparrow}(x)}\Bigg]\nonumber\\ &\times\Bigg[\uparrow\rightarrow\downarrow\Bigg] \nonumber \\
    =&a\int_x \frac{4}{\pi^2}\nabla\phi_{\uparrow}\nabla\phi_{\downarrow}+\frac{8}{(2\pi\alpha)^2}cos[2(\phi_{\uparrow}-\phi_{\downarrow})]\nonumber\\ &-\frac{8}{(2\pi\alpha)^2}cos[2(\phi_{\uparrow}+\phi_{\downarrow})]
\end{align}
The $cos4\phi_{\sigma}$ is highly irrelevant under RG compare to the cosines from (\ref{eq:Hm boson}) and therefore can be discarded without much concerns.

We now move on to terms coming from dimerisation $\mathcal{O}(\Delta)^1$. This amounts to looking for $e^{2ik_Fx}$ components from bosonisation as $(-1)^j=e^{2ik_Fx}$. Bosonising $H_0$ gives the following term
\begin{align}
     &-2B\Delta\sum_{\substack{j\\ \eta=\uparrow,\downarrow}}(-1)^j ( c^{\dagger}_{j+1,\eta}c_{j,\eta} + h. c. ) \nonumber \\
     &\approx \frac{16aB\Delta\pi}{(2\pi \alpha)^2} \sum_{\eta=\uparrow,\downarrow}\int_x sin2\phi_{\eta}
\end{align}
which is highly relevant. $H_{umk}$ requires some attention and bosonisation should be treated carefully within fermion normal ordering $\underline{\psi(x)_R\psi^{\dagger}(x')_R}=\left[2\pi(x-x')\right]^{-1} , \ \underline{\psi(x)_L\psi^{\dagger}(x')_L}=-\left[2\pi(x-x')\right]^{-1}$~\cite{orignac1998weakly}. In the end, this give $H_{umk}$ the following term
\begin{align}
    &-4\Gamma\Delta\sum_{\substack{j\\ \eta=\uparrow,\downarrow}}(-1)^j(c^{\dagger}_{j,\eta}c_{j,\eta}-\frac{1}{2})(c^{\dagger}_{j+1,\eta}c_{j+1,\eta}-\frac{1}{2})\nonumber \\
    =&\frac{-16a\Gamma\Delta}{(2\pi\alpha)^2}\sum_{\eta=\uparrow,\downarrow}\int_x sin2\phi_{\eta}(x)
\end{align}
and a less relevant operator $(\partial_x\phi)^2sin2\phi$ have been discarded. For $H_m$, the $2k_F$ component gives terms $~\nabla\phi_{\uparrow} cos2\phi_{\downarrow}+\nabla\phi_{\downarrow} cos2\phi_{\uparrow}$ which is irrelevant in the current model: By power counting, it can be seen that its dimension is $1+\frac{K_{\rho}+K_{\sigma}}{2}$. Since $K_{\rho},K_{\sigma}\geq1$ from (\ref{eq:Luttinger parameter}), this term is simply irrelevant in the current setting.

Inserting these results, and performing a unitary rotation to the charge and spin degree of freedom $\phi_{\rho}=\frac{\phi_{\uparrow}+\phi_{\downarrow}}{\sqrt{2}}, \ \phi_{\sigma}=\frac{\phi_{\uparrow}-\phi_{\downarrow}}{\sqrt{2}}$, we arrive at (\ref{eq:bosonised Hamiltonian}).

\section{RG flow for Sine-Gordon Hamiltonian}\label{sup:RG}
\begin{widetext}
The procedure here is a real space renormalisation group procedure that follows closely with Ref.~\onlinecite{giamarchi2003quantum,jose1977renormalization}. We will also demonstrate explicitly that the unmklapp term $H_{\mathrm{umk}}$ in (\ref{eq:effective spinful fermion Hamiltonian}) is way less relevant. The form of Sine-Gordon Hamiltonian we encounter from Umklapp term and dimerisation has the following form
\begin{align}
    H=&\sum_{i=1,2}\frac{1}{2\pi}\int dx \ u_iK_i(\partial_x\theta_i)^2+\frac{u_i}{K_i}(\partial_x\phi_i)^2 \nonumber \\&+ \frac{2g}{(2\pi\alpha)^2}\int dx\ cos(\beta\phi_1)cos(\beta\phi_2)
\end{align}
where $K_i,u_i$ are the Luttinger parameter and velocity of two different bosonic field species $\phi_{i},\theta_{i}$. $\beta$ is the frequency and it is $\sqrt{8}$ for the umklapp term while $\sqrt{2}$ for dimerisaiton term.
To begin with, consider the following correlation function
\begin{align}
    R(r_1-r_2)=\langle e^{ia^2\sqrt{2}\phi_1(r_1)}e^{-ia^2\sqrt{2}\phi_1(r_2)}\rangle_H
\label{average for renorm}
\end{align}
The average with respect to the free kinetic part of the Hamiltonian $H_0=\sum_{i=1,2}\frac{1}{2\pi}\int dx \ u_iK_i(\partial_x\theta_i)^2+\frac{u_i}{K_i}(\partial_x\phi_i)^2$ is 
\begin{align}
    \langle e^{ia^2\sqrt{2}\phi_i(r_1)}e^{-ia^2\sqrt{2}\phi_i(r_2)}\rangle_{H_0}=e^{-a^2K_iF_{1,i}(r_1-r_2)}\simeq(\frac{\alpha}{r_1-r_2})^{a^2K_i} \nonumber \\
    \langle[\phi(r_1)-\phi(r_2)]^2\rangle_{H_0}=K_iF_{1,i}(r_1-r_2) \ , \ F_{1,i}(r)=\frac{1}{2}\text{log}\left[\frac{x^2+(u_i|\tau|+\alpha)^2}{\alpha^2}\right]
\end{align}

Since the Hamiltonian is separable in the kinetic part, averages w.r.t. to the free kinetic Hamiltonian can be performed separably $\langle f(\phi_1)g(\phi_2)\rangle_{H_0}=\langle f(\phi_1)\rangle_{H_{0,1}}\langle g(\phi_2)\rangle_{H_{0,2}}$. The full action reads:
\begin{align}
    S=\overbrace{\sum_{i=1,2}\frac{1}{2\pi K_i}\int dxd\tau \frac{1}{u_i}(\partial_{\tau}\phi)^2+u_i(\partial_{x}\phi)^2}^{S_{0,1}+S_{0,2}} + \frac{2g}{(2\pi\alpha)^2}\int dxd\tau\ cos(\beta\phi_1)cos(\beta\phi_2)
\end{align}
$\theta$ has been integrated out as it merely contributes a constant which cancels out in the expectation value. As $u_1\neq u_2$, there is an extra non-trivial factor towards the end. If we expand in powers of $g$ the first order is 0 and stopping at second order, the partition function is
\begin{align}
    Z&=\int\mathcal{D}\phi_1\mathcal{D}\phi_2e^{-S} \nonumber \\
    &=\int\mathcal{D}\phi_1\mathcal{D}\phi_2e^{-S_{0,1}-S_{0,2}}\Bigg[1-0 \nonumber \\
    &+\frac{1}{32}\left(\frac{2g}{(2\pi\alpha)^2}\right)^2\int d^2r'd^2r''\prod_{i=1,2}\sum_{\epsilon_1,\epsilon_2=\pm}e^{i\epsilon_1\beta\phi_i(r')}e^{-i\epsilon_2\beta\phi_i(r')}\Bigg]
\end{align}
$d^2r=dxd\tau$ is different to the conventional definition for now. Expanding \ref{average for renorm} in $g$ and stopping at 2nd order, we have

\begin{align}
    &\langle e^{ia^2\sqrt{2}\phi_1(r_1)}e^{-ia^2\sqrt{2}\phi_1(r_2)}\rangle_H\approx e^{-a^2K_1F_{1,1}(r_1-r_2)} \nonumber \\
    &+\frac{1}{8}\left(\frac{g}{(2\pi\alpha)^2}\right)^2\Bigg[\int d^2r'd^2r''\langle e^{ia^2\sqrt{2}\phi_1(r_1)}e^{-ia^2\sqrt{2}\phi_1(r_2)}\prod_{i=1,2}\sum_{\epsilon_1,\epsilon_2=\pm}e^{i\epsilon_1\beta\phi_i(r')}e^{-i\epsilon_2\beta\phi_i(r')}\rangle_{H_0} \nonumber \\
    &-e^{-a^2K_1F_{1,1}(r_1-r_2)}\langle\prod_{i=1,2}\sum_{\epsilon_1,\epsilon_2=\pm}e^{i\epsilon_1\beta\phi_i(r')}e^{-i\epsilon_2\beta\phi_i(r')}\rangle_{H_0}\Bigg] \nonumber \\
    &=e^{-a^2K_1F_{1,1}(r_1-r_2)}\Bigg[1+\frac{1}{8}\left(\frac{g}{(2\pi\alpha)^2u_1}\right)^2\int d^2r'd^2r'' e^{-\frac{\beta^2}{2}\left(K_1F_{1,1}(r'-r'')+K_2F_{1,2}(r'-r'')\right)} \nonumber \\
    \times&2\sum_{\epsilon=\pm}\left(e^{\frac{a\beta}{\sqrt{2}}K_1\epsilon\left[F_{1,1}(r_1-r')-F_{1,1}(r_1-r'')+F_{1,1}(r_2-r'')-F_{1,1}(r_2-r')\right]}-1\right)\Bigg] \ , \ y=u_1\tau \nonumber \\
\label{renorm step 1}
\end{align}
Due to factor $e^{-\frac{\beta^2}{2}K_1F_{1,1}(r'-r'')}\sim(\frac{1}{r})^{\frac{\beta^2}{2}}$ which is a power law, only small $r'-r''$ contributes the most. Making 
\begin{align}
    R=\frac{r'+r''}{2}\ , \ r=r'-r'' \nonumber \\
    r_1-r'=r_1-R-\frac{1}{2}r \ , \ r_1-r''=r_1-R+\frac{1}{2}r
\end{align}
We can expand in $r$ giving
\begin{align}
    &\sum_{\epsilon=\pm}e^{\frac{a\beta}{\sqrt{2}}K_1\epsilon\left[F_{1,1}(r_1-r')-F_{1,1}(r_1-r'')+F_{1,1}(r_2-r'')-F_{1,1}(r_2-r')\right]}-1 \nonumber \\
    \approx&\frac{a^2\beta^2}{2}K_1^2\left[\sum_{i,j=x,y}r_i\nabla_{R_j}\left(F_{1,1}(r_1-R)-F_{1,1}(r_2-R)\right)\right]^2
\end{align}
the integral is only non-zero for $i=j$ ($i\neq j$ odd function) and $\int d^2r x^2=\int d^2r y^2=\int d^2r \frac{r^2}{2}$. With integration by part \ref{renorm step 1} becomes,

\begin{align}
    &=e^{-a^2K_1F_{1,1}(r_1-r_2)}\Bigg[1-\frac{1}{16}\left(\frac{g}{(2\pi\alpha)^2u_1}\right)^2\int d^2rd^2R\ e^{-\frac{\beta^2}{2}\left(K_1F_{1,1}(r)+K_2F_{1,2}(r)\right)} \nonumber \\
    \times&a^2\beta^2K_1^2r^2\left[F_{1,1}(r_1-R)-F_{1,1}(r_2-R)\right]\left(\nabla^2_X+\nabla^2_Y\right)\left[F_{1,1}(r_1-R)-F_{1,1}(r_2-R)\right] \nonumber \\
\end{align}
Since $F_{1,1}(r)\simeq\text{log}(\frac{r}{\alpha})$ for $r>\alpha$, we can use the following identity
\begin{align}
    \left(\nabla^2_X+\nabla^2_Y\right)\text{log}(R)=2\pi\delta(R)
\end{align}
and $\int d^2R \left[F_{1,1}(r_1-R)-F_{1,1}(r_2-R)\right]\left(\nabla^2_X+\nabla^2_Y\right)\left[F_{1,1}(r_1-R)-F_{1,1}(r_2-R)\right]=-4\pi F_{1,1}(r_1-r_2)$. $F_{1,1}(0)=0$ with regularisation. At this point we need to remember that
\begin{align}
    F_{1,2}(r)=\text{log}\left[\frac{\sqrt{x^2+(\frac{u_2}{u_1}|y|+\alpha)^2}}{\alpha}\right]\simeq\text{log}\left[\frac{\sqrt{x^2+(\frac{u_2}{u_1}|y|)^2}}{\alpha}\right]=\text{log}\left[\frac{r\sqrt{1+\epsilon_1 cos^2(\theta)}}{\alpha}\right]
\end{align}
$\left(\frac{u_2}{u_1}\right)^2=1+\epsilon_1$, and the above behaviour for $F_{1,2}$ approximately holds true provided $\epsilon>-1$ (regularisation can be appropriately ignored). \ref{average for renorm} is thus
\begin{align}
    R(r_1-r_2)\approx &e^{-a^2K_1F_{1,1}(r_1-r_2)}\Bigg[1+F_{1,1}(r_1-r_2)\frac{a^2\beta^2K_1^2\pi}{4}\left(\frac{g}{(2\pi\alpha)^2u_1}\right)^2\int_{r>\alpha} d^2r\ e^{-\frac{\beta^2}{2}\left(K_1F_{1,1}(r)+K_2F_{1,2}(r)\right)}r^2 \Bigg] \nonumber \\
    =e^{-a^2K_1F_{1,1}(r_1-r_2)}&\Bigg[1+F_{1,1}(r_1-r_2)\frac{a^2\beta^2K_1^2}{2\pi}\frac{g^2}{32\pi^2\alpha^4u_1^2}\int_{\alpha}^{\infty} r^3dr\int_{-\pi}^{\pi}d\theta\ (\frac{\alpha}{r})^{\frac{\beta^2}{2}(K_1+K_2)}(\frac{1}{1+\epsilon_1cos\theta})^{\frac{\beta^2K_2}{4}} \Bigg] \nonumber \\
    =e^{-a^2K_1F_{1,1}(r_1-r_2)}&\Bigg[1+F_{1,1}(r_1-r_2)\frac{a^2\beta^2K_1^2}{32}\frac{\tilde{g}^2I(\epsilon_1,K_2,\beta)}{2\pi}\int_{\alpha}^{\infty}\left(\frac{\alpha}{r}\right)^{\frac{\beta^2}{2}(K_1+K_2)-3}\frac{d r}{\alpha} \Bigg]
\end{align}

$\tilde{g}=\frac{g}{\pi u_1}$,$I(\epsilon_1,K_2,\beta)=\int_{-\pi}^{\pi}d\theta (\frac{1}{1+\epsilon_1cos\theta})^{\frac{\beta^2K_2}{4}}$. The bracket can be re-exponentialised giving.
\begin{align}
    K_{1,eff}(\alpha)=K_1-\frac{g(\alpha)^2\beta^2K_1^2}{32\pi^2u_1^2}\frac{I(\epsilon_1,K_2,\beta)}{2\pi}\int_{\alpha}^{\infty}\left(\frac{\alpha}{r}\right)^{\frac{\beta^2}{2}(K_1+K_2)-3}\frac{d r}{\alpha}
\end{align}
If we change the cutoff $\alpha=\alpha'+d\alpha$ and re-parametrise $\alpha=\alpha_0e^l$, $K_{1,eff}$ and $g$ has to change accordingly giving the renormalisation group flow shown below. The flow for $K_2$ can be worked out with the same procedure but replacing $\phi_1\rightarrow\phi_2$ in the correlator $\langle e^{ia^2\sqrt{2}\phi_i(r_1)}e^{-ia^2\sqrt{2}\phi_i(r_2)}\rangle_{H_0}$. All in all, we have
\begin{align}
    \partial_lK_1=-\frac{g^2\beta^2K_1^2}{32\pi^2u_1^2}\frac{I(\epsilon_1,K_2,\beta)}{2\pi} \nonumber \\
    \partial_lK_2=-\frac{g^2\beta^2K_2^2}{32\pi^2u_2^2}\frac{I(\epsilon_2,K_1,\beta)}{2\pi} \nonumber \\
    \partial_lg=\Bigg(2-\frac{\beta^2}{4}(K_1+K_2)\Bigg)g
\label{RG CosCos}
\end{align}
$\epsilon_2=-\frac{\epsilon_1}{1+\epsilon_1}$. With this we can immediately tell the Umklapp term $\beta=\sqrt{8}$ is simply less relevant while for dimerisation $\beta=\sqrt{2}$ is highly relevant.

The RG flow for usual Sine-Gordon Hamiltonian of the form $H=\frac{1}{2\pi}\int dx \ uK(\partial_x\theta)^2+\frac{u}{K}(\partial_x\phi)^2 + \frac{2g}{(2\pi\alpha)^2}\int dx\ cos(\beta\phi)$ can be worked out similarly and the extra factor $I(\epsilon_1,K_2,\beta)$ reduces to 1.
\end{widetext}

\section{Details about numerics}\label{app:numerics}

The procedures for our numerics employed follow the steps described in~\cite{kells2023topological}, which is an extension of~\cite{kells2023topological} to generic particle non-conserving case. Across all simulation, the length of the associated complex fermion chain is set to multiple of 4 and we employ open boundary condition in order to compute a meaningful topological entanglement entropy. The discrete time parameter $\delta t$ has been chosen to be $0.05$, and the number of trajectories for each set of parameters are typically above 600.

To simulate the PPS dynamics, we employ (\ref{eq:PPS SSE}), assuming inhomogeneous measurement strength, and since the measurement operators $\hat{O}_j$'s square to $\mathbb{I}$, it reduces to
\begin{align}\label{eq:PPS SSE for numerics}
    d\ket{\psi_{t} } &= \frac{1}{N}\bigg[(-iH - \sum_j\gamma_j\hat{O}_j\langle \hat{O}_j\rangle+ \sum_jB_j \hat{O}_j)dt \nonumber \\ &+\sum_j dW_j \hat{O}_j\bigg]\ket{\psi_t}
\end{align}
where $N$ is some normalisation, $dW_jdW_j'=\gamma_jdt\delta_{j,j'}$ is the Weiner process and we have absorbed any operator independent term into the normalisation. One can exponentialise this expression giving
\begin{align}
    \ket{\psi_{t+dt}}&=\frac{1}{N_1}\exp\bigg[-iHdt-dt\sum_j\gamma_j\hat{O}_j\langle \hat{O}_j\rangle+dt\sum_jB_j \hat{O}_j \nonumber \\ &+\sum_j dW_j \hat{O}_j\bigg]\ket{\psi_t}
\end{align}
As $H$ is a white noise, with homogeneous strength and share the same set of operators $\hat{O}_j$ with measurement, this is further modified to
\begin{align}\label{eq:exp PPS SSE}
    \ket{\psi_{t+dt}}&=\frac{1}{N_1}\exp\bigg[-i\sum_j\hat{O}_jd\xi_j- dt\sum_j\gamma_j\hat{O}_j\langle \hat{O}_j\rangle\nonumber \\ &+dt\sum_jB_j \hat{O}_j +\sum_j dW_j \hat{O}_j\bigg]\ket{\psi_t}
\end{align}
where $d\xi_j\xi_j'=J^2\delta_{j,j'}dt,d\xi_jdW_j'=0$ is another Weiner process. In the case of deterministic unitary, one can trotterise the update into measurement and unitary separately 
\begin{align}
    \ket{\psi_{t+\delta t}}=e^{-iH\delta t}e^{- \delta t\sum_j\gamma_j\hat{O}_j\langle \hat{O}_j\rangle\nonumber+\delta t\sum_jB_j \hat{O}_j +\sum_j \delta W_j \hat{O}_j}\ket{\psi_t}
\end{align}
where $\delta t$ is the discrete time interval and $\delta W_j$'s are random variables with mean 0 and variance $\gamma\delta t$. In the lowest order of error, it is merely $\comm{H\delta t}{\delta W_j\hat{O}_j}\sim\mathcal{O}(\delta t^{3/2})$ which vanishes as $\delta t$ is reduced. (\ref{eq:exp PPS SSE}) however includes white noises as unitary update and lowest order of error becomes $\mathcal{O}(\delta t)$ which does not vanishes in the time continuum limit. Therefore, the safest route is to not trotterise the update into measurement and unitary blocks, and instead retain them in a single exponential while discretising time with $\delta t$, $\delta W_j$ and $\delta \xi_j$ (mean 0 and variance $J^2\delta t$). We have numerically check that this does have a slight effect on the outcome of the simulation.

To simulate the Majorana chain, we implement the calculate in the Bogoliubov de Gennes
(BdG) formalism, by first identifying 1 species of complex fermion to rewrite the chain: $c^{\dagger}_j=(\chi_{2j-1}+i\chi_{2j})/2$. The operators of interest, which are the odd and even bond parity, become the on-site and cross-site parity:
\begin{align}\label{eq:complex ferm sim}
    i\chi_{2j-1}\chi_{2j}&=(1-2c^{\dagger}_jc_j)\nonumber \\ i\chi_{2j}\chi_{2j+1}&=(c^{\dagger}_j-c_j)(c^{\dagger}_{j+1}+c_{j+1})
\end{align}
As the model including the measurement is Gaussian preserving, starting from a Guassian state, the evolution will remain in the space of Gaussian states.  For a generic Guassian state, one can express it as~\cite{mbeng2020quantum,kells2023topological}
\begin{align}\label{eq:Gaussian state}
    \ket{\psi}=\prod_{n=1}^{n=L}\sum_{k,n}V^*_{k,n}c_k^{\dagger}+U^*_{k,n}c_k
\end{align}
where $V$ and $U$ are $L\times L$ matrices which form a $2L\times2L$ orthonormal matrix
\begin{align}\label{eq:W matrix}
    W=\begin{pmatrix}
        U &\ V^* \\
        V &\ U^*
    \end{pmatrix}
\end{align}
and implies $U^{\dagger}U+V^{\dagger}V=\mathbb{I},U^TV+V^TU=0$. Any Gaussian state is fully characterised by the set of all two point correlators; All two point correlation can be calculated from directly from V and U as
\begin{align}\label{eq:2 point}
    C_{i,j}=\langle c^{\dagger}_ic_j\rangle= V^*V^{T}\nonumber \\
    F_{i,j}=\langle c_ic_j\rangle=V*U^{T}
\end{align}
Therefore, it is enough to evolve the matrices $U$ and $V$ alone. To achieve this, the white noise and measurement are written in the basis of complex fermion shown in (\ref{eq:complex ferm sim}) giving 2 separate non-commuting set of white noise and measurement. In the BdG formalism, each set of noise/measurement is represented by a matrix:
\begin{align}
    \sum_j(1-2c^{\dagger}_jc_j)\equiv\underline{c}^{\dagger}M_{2j-1}\underline{c} \nonumber \\
    \sum_j (c^{\dagger}_j-c_j)(c^{\dagger}_{j+1}+c_{j+1}) \equiv\underline{c}^{\dagger}M_{2j}\underline{c}
\end{align}
where $\underline{c}=(c^{\dagger}_1,c^{\dagger}_2,\dots,c^{\dagger}_L,c_1,\dots,c_L)^T$, and the matrices are
\begin{align}\label{eq:meas noise BdG}
    M_{2j-1}&=2\mathbb{I}_{L\times L} \nonumber \\
    M_{2j}&=\begin{pmatrix}
        -A &\ B^{\dagger} \\
        B &\ A
    \end{pmatrix} \nonumber \\
   A=diag(1,1)+&diag(1,-1) \ , B=-diag(1,1)+diag(1,-1)
\end{align}
$diag(1,\pm1)$ indicate 1 along the $\pm1$ off diagonal. Dimerisation implemented in the original Majorana chain corresponds to grouping the measurement strengths into 2 sets $\{\gamma_{2j-1}\}=\gamma,\{\gamma_{2j}\}=\alpha$, each with uniform strength within them (and hence $B_j$ into $B_{\gamma},B_{\alpha}$) and the ratio gives the dimerisation $\frac{1-\Delta}{1+\Delta}=\frac{\gamma}{\alpha}$. Denoting $(1-2c^{\dagger}_jc_j)=\hat{\Gamma}_j$ and $(c^{\dagger}_j-c_j)(c^{\dagger}_{j+1}+c_{j+1})=\hat{A}_j$, (\ref{eq:exp PPS SSE}) becomes
\begin{widetext}
\begin{align}\label{eq:ready for BdG}
    \frac{1}{N}\exp\bigg[&-i\sum_j\hat{\Gamma}_jd\xi_{1,j}-i\sum_j\hat{A}_jd\xi_{2,j} -\gamma dt\sum_j\hat{\Gamma}_j\langle\hat{\Gamma}_j\rangle -\alpha dt\sum_j\hat{A}_j\langle\hat{A}_j\rangle +B_{\gamma}dt\sum_j\hat{\Gamma}_j+B_{\alpha}\sum_j\hat{A}_j\nonumber \\ &+\sum_j\hat{\Gamma}_jdW_{\gamma,j}+\sum_j\hat{A}_jdW_{\alpha,j}\
    \bigg]
\end{align}
\end{widetext}
$d\xi_{k,j}d\xi_{l,j'}=J^2dt\delta_{j,j'}\delta_{k,l}$, $dW_{\gamma,j}dW_{\gamma,j'}=\gamma dt\delta_{j,j'}$ and $dW_{\alpha,j}dW_{\alpha,j'}=\alpha dt\delta_{j,j'}$. The update of the matrices $V$ and $U$ can now be implemented in the BdG form, and in the first step, they are multiplied by:
\begin{align}
    \begin{pmatrix}
        \tilde{U}(t+\delta t) \\
        \tilde{V}(t+\delta t)
    \end{pmatrix}=\exp[M]\begin{pmatrix}
        \tilde{U}(t) \\
        \tilde{V}(t)
    \end{pmatrix}
\end{align}
where $M$, a matrix, is merely the exponential in (\ref{eq:ready for BdG}) written in BdG form, and the operators are replaced by matrices of the form in (\ref{eq:meas noise BdG}) where entries are appropriately multiplied by the random variables $\delta\xi_{k,j}$ , $\delta W_{\gamma,j}$ and $\delta W_{\alpha,j}$ . The expectation values present can readily be computed from two point correlators in and (\ref{eq:2 point}). As $\tilde{U}$ and $\tilde{V}$ does not meet the criterion below (\ref{eq:W matrix}), a final step involves a normalisation of the state to ensure $W$ is orthonormal, which can be implemented via any orthonoramlisation procedure of a matrix: QR, Gram-Schmidt or singular value decomposition. Here, we chose QR and the final update is
\begin{align}
    QR=\begin{pmatrix}
        \tilde{U}(t+\delta t) \\
        \tilde{V}(t+\delta t)
    \end{pmatrix} \ , \begin{pmatrix}
        U(t+\delta t) \\
        V(t+\delta t)
    \end{pmatrix}=Q
\end{align}
$U(t+\delta t)$ and $V(t+\delta t)$ are now properly normalised.

To compute the entanglement entropy, recall that the Nambu one-body Green's function matrix is
\begin{align}\label{eq:Nambu Green}
    G=\begin{pmatrix}
        \mathbb{I}_{L\times L}-C^T &\ F \\
        F^{\dagger} &\ C
    \end{pmatrix}
\end{align}
The entanglement entropy of a subsystem $\mathbf{A}$ is calculated by reducing the Green's function to only fermions operators in $\mathbf{A}$, $G_{\mathbf{A}}$, and is given by~\cite{alba2009entanglement}
\begin{align}
    S_{1,\mathbf{A}}=-\sum_{\{\lambda_j\}}[\lambda_jlog_2\lambda_j+(1-\lambda_j)log_2\lambda_j]
\end{align}
where $\{\lambda_j\}$ are the set of eigenvalues of $G_{\mathbf{A}}$. For completeness, higher order entropies are
 \begin{align}
     S_{n,\mathbf{A}}=\frac{1}{1-n}\sum_{\{\lambda_j\}}log_2\left[\left(\lambda_j\right)^n+\left(1-\lambda_j\right)^n\right]
 \end{align}

\begin{figure}
\includegraphics[width=0.45\textwidth]{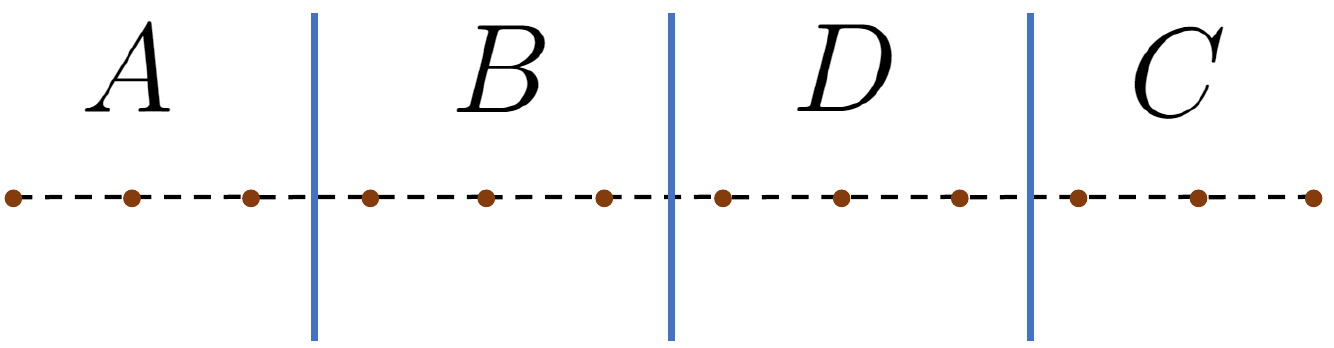}
\caption{Partition of the system for topological entanglement entropy calculation. The blue lines divide the system into 4 equal partitions.}
\label{fig:TEE partition}
\end{figure}

 To extract the critical exponenet $\nu$ in the measurement-only scenario, the topological entanglement entropy ($S_{TEE}$) for 1D system is needed and it is computed as~\cite{zeng2016topological}:
 \begin{align}
     S_{TEE}=S_{AB}+S_{BC}-S_{B}-S_{ABC}
 \end{align}
 where the partition $A$, $B$ and $C$ is pictured in figure \ref{fig:TEE partition}.
It is convenient to fix the on-site parity measurement strength $\gamma=1$, and one need only to adjust $\alpha$, which in the thermodynamic limit the critical point is located at $\alpha_{crit}=1$ (in practice, due to the length being even, there is always one less cross-site measurement and $\alpha_{crit}$ will be deviated slightly from 1). $S_{TEE}$ is computed for different system size $L$ across a range of $\alpha$ in the vicinity of $\alpha=1$, the expected critical point, at a given $b$ (see blow (\ref{eq:PPS SSE})). Near the critical point, $S_{TEE}$ is expected to have the following scaling form~\cite{sandvik2010computational,lavasani2021topological,kells2023topological}:
\begin{align}\label{eq:TEE scaling form}
    S_{TEE}=F\left[(\alpha-\alpha_{crit})L^{\frac{1}{\nu}}\right],
\end{align}
and $\alpha_{crit}$ the true numerical critical point can be read off as the crossing point of $S_{TEE}$ from various $L$'s.
%\begin{figure}
%\includegraphics[width=0.45\textwidth]{paper writing/supplemental/numerics/TEE show.pdf}
%\caption{A typical figure for data collapse of the topological entanglement entropy, using the function \eqref{eq:TEE scaling form}. The inset shows the plot of the optimising function \eqref{eq:optimising function nu} for various $\nu$ at a given $\alpha_{crit}$.}
%\label{fig:TEE crossing}
%\end{figure}
and is generally found to be $1.00\pm0.01$ (in practice, it may appear over a range but it is narrowed down further which we explain below). Although one notable exception is $b=1.5$, $\alpha_{crit,b=1.5}=0.975$, appears to deviate more, it is still with the $5\%$ error. $\nu$ is used as a fitting parameter for the data collapse, and its value is determined by the `best' data collpase which is quantified by the following objective function~\cite{lavasani2021topological}:
\begin{align}\label{eq:optimising function nu}
    \epsilon(\nu)&=\sum_{i=2}^{n-1}(y_i-\overline{y}_i)^2 \nonumber\\
    \text{where} \ \overline{y}_i&=\frac{(x_{i+1}-x_{i})y_{i-1}-(x_{i-1}-x_{i})y_{i+1}}{x_{i+1}-x_{i-1}}
\end{align}
$x_i$ are defined to be $(\alpha_i-\alpha_{crit})L^{1/\nu}$ and $y_i=S_{TEE}(\alpha_i,L_i)$. $i$ labels different data points and its ordering is sorted on the basis of ascending order in $x_i's:x_1<x_2<\dots x_n$. The `best' data collapse corresponds to the minimum of $\epsilon(\nu)$, at a given $\alpha_{crit}$, and we follow the convention in~\cite{lavasani2021measurement,lavasani2021topological} to define the error as the range of $\nu$ which falls within 2 times the minimum $\epsilon(\nu)<2\epsilon(\nu)_{min}$. In addition, $\alpha_{crit}$ is further narrowed down by locating the global minimum of $\epsilon(\nu)$, accounting for $\alpha_{crit}$ as well.

As a final point, to distinguish numerically $(\text{log}L)^2$ from $(\text{log}L)$, one should not employ the fit $S_0\propto(\text{log}L)^n$ as sub-leading term could overshadow the scaling. Instead, one should employ the difference~\cite{fava2023nonlinear}
\begin{align}\label{eq:EE diff fit log or log 2}
    \delta S_{0,L}=S_{0,2L/2}-S_{0,L/2}
\end{align}
where $S_{0,L/2}$ is the half-system entanglement entropy. The subleading term are therefore cancelled in $\delta S_{0,L}$, and the scaling are different:
\begin{align}\label{eq:EE diff scale}
    \delta S_{0,L}\propto\begin{cases}
        \text{log}_2L, \ \text{if} \ S_{0,L/2}\sim (\text{log}\frac{L}{2})^2 \\
        \text{constant}, \ \text{if} \ S_{0,L/2}\sim \text{log}\frac{L}{2}
    \end{cases}
\end{align}

\begin{figure}
\includegraphics[width=0.45\textwidth]{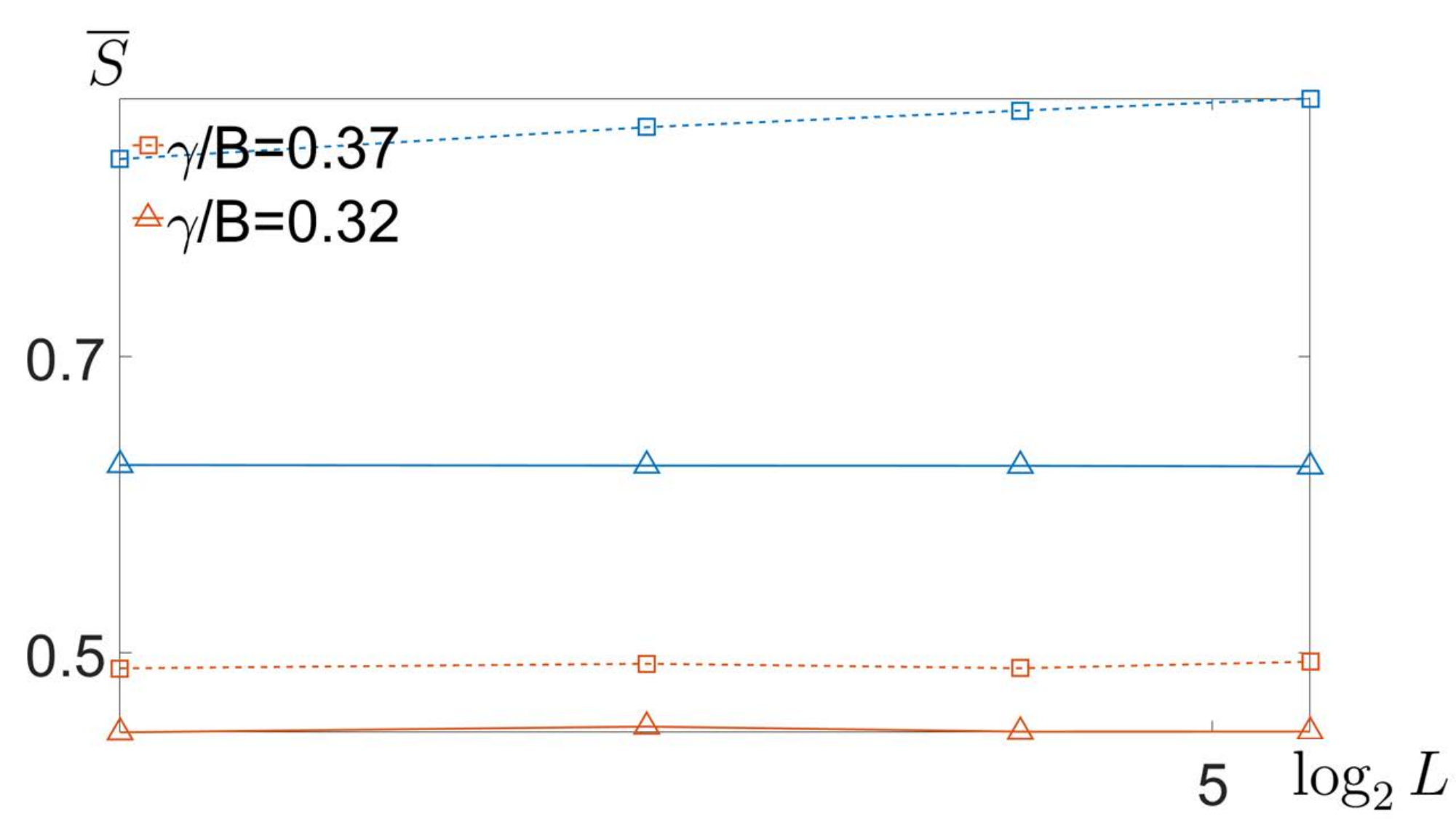}
\caption{ 
Average entanglement entropy $\overline{S}_0$  from numerical simulations to demonstrate the area law and logarithmic entanglement scaling phases for non-zero dimerization. The two set of lines, dashed square and solid triangle, corresponds to the different PPS strength. The two different colours within each set correspond to the orange and blue points in Fig.~\ref{fig:phases from numeric and 2 replica cal for general}
}
\label{fig:phases from numerics for general}
\end{figure}

We also show in Fig.~\ref{fig:phases from numerics for general} the numerical data supporting Fig.~\ref{fig:phases from numeric and 2 replica cal for general} from analytical calculation.

\pagebreak
\bibliography{bib.bib}

\end{document}